\documentclass[aps,prd,superscriptaddress,showpacs,tighten,nofootinbib,twocolumn]{revtex4}
\usepackage{microtype,cmap,amssymb}
\usepackage{graphicx}
\usepackage{float}
\usepackage{subfig}
\usepackage{epsfig}
\usepackage{amsmath}
\usepackage{fixmath}
\newcommand{\bea}{\begin{eqnarray*}}
\newcommand{\eea}{\end{eqnarray*}}

\begin{document}


\title{Non-standard interaction in neutrino oscillations  and recent Daya Bay, T2K experiments}

\author{Rathin Adhikari}
\email{rathin@ctp-jamia.res.in}
\affiliation{Centre for Theoretical Physics, Jamia Millia Islamia 
(Central University),  Jamia Nagar, New Delhi-110025, INDIA}

\author{Sabyasachi Chakraborty}
\email{tpsc3@iacs.res.in}
\affiliation{Department of Theoretical Physics, Indian Association for 
the Cultivation of Science, 2A $\&$ 2B Raja S.C.Mullick Road, Jadavpur, 
Kolkata 700 032, INDIA}

\author{Arnab Dasgupta}
\email{arnab@ctp-jamia.res.in}
\affiliation{Centre for Theoretical Physics, Jamia Millia Islamia 
(Central University), Jamia Nagar, New Delhi-110025, INDIA}

\author{Sourov Roy}
\email{tpsr@iacs.res.in}
\affiliation{Department of Theoretical Physics, Indian Association for 
the Cultivation of Science, 2A $\&$ 2B Raja S.C.Mullick Road, Jadavpur, 
Kolkata 700 032, INDIA}

\begin{abstract}

We study the possible constraints on non-standard interaction(NSIs) in a model
independent way by considering the recent results from T2K and Daya Bay neutrino 
oscillations experiments. Using perturbation method we present generic formulas 
(suitable for T2K baseline and for large $\theta_{13}$ as evident from Daya Bay) 
for the probability of oscillation for $\nu_\mu \rightarrow \nu_e$, taking into 
account NSIs at source ($\epsilon^s$), detector ($\epsilon^d$) and during propagation 
($\epsilon^m$) of neutrinos through matter. Two separate cases of perturbation with 
small (slightly large) NSI ($\epsilon_{\alpha\beta}^m \sim 0.03(0.18)$) are discussed in detail.
Using various possible presently allowed NSI values we  reanalyze numerically 
the $\theta_{13} - \delta$ allowed region given by recent T2K experimental data. We 
obtain model independent constraints on NSIs in the $\delta-\epsilon_{\alpha\beta}^m$ 
plane using the $\theta_{13}$ value as measured by Daya Bay, where $\delta$ is the 
CP violating phase. Depending on $\delta$ values significant constraints on 
$\epsilon_{e\tau}$ and $\epsilon_{\tau\tau}$, in particular, are possible for both 
hierarchies of neutrino masses. Corresponding to T2K's 66\% confidence level result, 
the constraints on $\epsilon_{\tau\tau}$ is shown to be independent of any $\delta$ value.
\end{abstract}

\maketitle
\section{Introduction}
Neutrino oscillations successfully describe neutrino flavor transitions. The recent superbeam 
and reactor neutrino experiments have provided enormous insights 
to unravel the exact value of the vacuum mixing angle $\theta_{13}$. To emphasize this point, 
the T2K\cite{T2K} experiment observed indications of $\nu_{\mu}\rightarrow\nu_{e}$ 
appearance, by producing a conventional neutrino beam at J-PARC and directed 
$2.5^{\circ}$ off-axis to a detector situating at $295$ Km away. The bounds on 
$\theta_{13}$ which, T2K came up with was $0.03(0.04)<\sin^2 2\theta_{13}< 0.28(0.34)$ 
for $\delta=0$ and normal (inverted) hierarchy. The reactor neutrino experiments 
like Daya Bay\cite{Daya Bay} and Reno\cite{Reno} provided compelling evidences for a 
relatively large angle $\theta_{13}$, with $5.2\sigma$ and $4.9\sigma$ results 
respectively. These recent reactor neutrino results indicate $\theta_{13}$ very 
close to $8.8^{\circ}$.\\

In this work we considered non-standard interactions (NSIs), occurring from four-fermion 
operators. In addition to the standard model Lagrangian density, we consider the following 
non-standard interactions in the low energy effective theory during the propagation of 
neutrinos through matter
\begin{eqnarray}
\mathcal{L}^M_{NSI} &= -2\sqrt{2}G_F\epsilon^{f P}_{\alpha \beta}\left[\bar{f}\gamma^{\mu}P 
f\right]\left[\bar{\nu}_{\alpha}\gamma_{\mu}P_L\nu_{\beta}\right],
\end{eqnarray}
where $f=e,u,d$ and $P=P_L,P_R$ where $P_L=(1-\gamma_5)/2$ and $P_R= (1+ \gamma_5)/2$.
 In our subsequent sections these non-standard interactions (NSIs) are related to 
$\epsilon_{\alpha\beta}^m$ as mentioned later. In the neutrino oscillation experiments 
for the NSIs at the source and detector the following lagrangian densities as low energy 
effective theory corresponding  to charged current interactions due to leptons and quarks 
may be considered: 
\begin{eqnarray}
\mathcal{L}^{S,D}_{NSI}=-2\sqrt{2}G_F\epsilon^{\lambda \sigma P}_{\gamma \delta}
\left[\bar{l}_{\lambda}\gamma^{\mu}P l_{\sigma}\right]\left[\bar{\nu}_{\gamma}
\gamma_{\mu}P_L\nu_{\delta}\right] 
\end{eqnarray}
\begin{eqnarray}
\mathcal{L}^{S,D}_{NSI}=-2\sqrt{2}G_F\epsilon^{udP}_{\gamma \delta}V_{u d}
\left[\bar{u}\gamma^{\mu}P d\right]\left[\bar{l}_{\gamma}\gamma_{\mu}P_L
\nu_{\delta}\right]\nonumber \\ + h.c
\end{eqnarray}
In our subsequent sections both these NSIs - $\epsilon^{\lambda \sigma P}_{\gamma \delta}$ 
and $ \epsilon^{udP}_{\gamma \delta}$ contribute to $\epsilon_{\alpha\beta}^s$ and 
$\epsilon_{\alpha\beta}^d$ corresponding to the appropriate interactions at source and 
detector for the neutrinos respectively. From the Lagrangian, we observe that the NSI 
parameters do not possess any mass dimension. However, if NSIs are related to the underlying 
new physics, then they should be considered as a first order term in the perturbation series 
and not in the zeroth order\cite{Meloni}. To reiterate this, we know the NSI parameters
are related to the new physics scale in the form $\epsilon\sim(M_{W}/M_{NSI})^2$,
where $M_{NSI}$ signifies the new physics scale. So if we consider the new physics
scale to be around a few TeV, then the NSI parameters should not be greater
than a few percent. In general the NSI parameters can be categorized into two 
different parts. One is the NSI during propagation, and the other being NSI at the 
source and at the detector. It is worthwhile to note that the present bounds on the 
NSI parameters during propagation are not very stringent\cite{genbound-nsi}. \\

Non Standard interaction and its implications in a model independent way, as well 
as in different models have already been studied very extensively in the literature  
\cite{Grossman,Ota,Huber,Ota1,GonzalezGarcia,Gago,Huber1,Huber2,Campanelli,Davidson1,
Honda,Kitazawa,Blennow}. Many authors have studied their impact on solar neutrinos 
\cite{Bergmann,Berezhiani,Friedland,Miranda}, atmospheric neutrinos 
\cite{GonzalezGarcia1,Bergmann1,Fornengo,GonzalezGarcia2,Friedland2,Bergmann2}, 
conventional and upgraded neutrino beams\cite{Ota,Ota1,Takeuchi,Kitazawa,
Friedland:2006pi,Blennow}, neutrino factories\cite{Ota,GonzalezGarcia,Huber1,Gago,Huber,
Campanelli,Bueno,Kopp:2007mi,winter}, beta beams\cite{Rathin}, supernova neutrinos 
\cite{Fogli,Duan,Esteban}, cosmological relic neutrinos\cite{Mangano} and 
neutrino-instability problem\cite{khlopov}. Also in some cases similarity of such 
effective interactions with CPT violations\cite{diaz} and probing such interactions 
at LHC \cite{shoemaker} have been discussed. In the context of solar neutrinos 
possible confusion of nonzero mixing angle $\theta_{13}$ in the presence of NSI and 
hint of NSI have been mentioned \cite{palazzo}. One of the most striking features 
of NSI parameters, is to cloud the sensitivity of $\theta_{13}$ by orders of magnitude, 
which was shown very explicitly in \cite{Huber2,Zhang}, for neutrino factories and 
reactor neutrino experiments respectively. \\

To elaborate the plan of our paper, at first in section II, we present a generalized 
prescription (suitable for relatively short baseline of T2K) by following
 the works of \cite{Minakata,Minakata1,Minakata2}, which in 
the literature is also known as the `$\sqrt{\epsilon}$ method of perturbation theory', 
where $\epsilon\equiv\frac{\Delta m^2_{21}}{\Delta m^2_{31}}\sim 0.03$ and represent 
a mathematical formulation by considering a relatively large $\sin\theta_{13}\sim\sqrt
{\epsilon}\sim0.18$. We divide the Hamiltonian, consisting of the standard matter 
interaction and NSI during propagation, into zeroth order part and a perturbative 
part, where $\sqrt{\epsilon}$ is the perturbation parameter. Our next task is to 
compute the S-matrix elements from these Hamiltonians. In section III, we invoke the 
idea of NSI parameters at the source and the detector. Previous bounds on these 
parameters were constrained by lepton and pion decays\cite{Bergmann1,GonzalezGarcia}, 
which were of the order of $\mathcal{O}(0.1)$. However, the present bounds on NSI 
parameters at the source and at the detector are very strong\cite{genbound-nsi}. 
Due to this reason we assume the NSI parameters, present at the source and the detector 
are of the order of $\epsilon$. In section IV and V we have considered two different 
cases, one with the consideration of $\epsilon_{\alpha\beta}^m\sim\sqrt{\epsilon}$ and the other 
case with $\epsilon_{\alpha\beta}^m\sim\epsilon$, where $\epsilon_{\alpha\beta}^m$ is the 
NSI parameter during propagation. We have presented the expression of the probability up 
to second order in $\epsilon$, by taking into consideration all these effects, such as the standard 
matter interaction, NSI during propagation and NSI at the source and at the detector.
We were able to match the results obtained from the analytical expressions with that of the full 
numerical study. This also shows the remarkable power of this 
perturbation method. We show that because of the presence of NSIs at the source and 
at the detector, one can have a non-zero oscillation probability at the source itself 
without the neutrino traversing any length. This is coined as the zero distance effect
\cite{Yasuda,antus,London,Rodejohann} or the near detector effect\cite{Kopp} in the 
literature. This effect is a manifestation of the  non-unitarity of the mixing matrix 
by considering NSIs at the source and at the detector.

It is also important to note that, in principle one can also follow the method of
matrix perturbation, to obtain the expression for the probability. In that case one has 
to compute the modification of the PMNS matrix\cite{Pontecorvo,Maki}, due to the 
inclusion of the standard matter interaction and NSI during propagation. The modified 
PMNS matrix has to be diagonalised, the eigenvectors and the eigenvalues are to be 
extracted from the modified PMNS matrix. After that NSI at the source and at the detector 
are to be included, to compute the overall expression of the oscillation probability. 
Similar approaches were followed by the authors of\cite{Akhmedov,Freund} 

In   section VI, using Daya Bay and T2K experimental results we have discussed numerical analysis in obtaining the
 constraints in the $\delta$-NSI plane. Here, the larger model-independent allowed values of NSI (not considered in 
 our perturbative approach) have been considered for the analysis.

\section{\bf{Mathematical Formulation for Large $\theta_{13}$ 
Perturbation Theory}}

The recent reactor based neutrino experiments have provided substantial proof for a 
relatively larger $\theta_{13}$. Based on the works of \cite{Minakata,Minakata1,Minakata2}, 
we describe a mathematical prescription to show the effects of the non-standard 
interactions during propagation in neutrino oscillations. We consider the channel 
$\nu_{\mu}\rightarrow\nu_e$, as followed by the recently concluded T2K experiment. 
Using the present experimental values of 
$\theta_{13}$ and the mass squared differences, we formulate
\begin{eqnarray} 
\sin\theta_{13}=s_{13} \sim \sqrt{ \epsilon }, 
\hspace{8mm} 
\epsilon  \equiv \frac{ \Delta m^2_{21} }{ \Delta m^2_{31} }  \sim 0.03.
\label{ansatz}
\end{eqnarray}
This section elaborates the basic principles of our perturbative approach. In the 
Shr\"{o}dinger picture, a neutrino with flavor $\alpha$ obeys the evolution 
equation\cite{Boris}
\begin{equation}
i \frac{d |\nu_{\alpha}(t) \rangle}{dt} = \mathcal{H}|\nu_{\alpha}(t) 
\rangle; \hspace{1.5cm} |\nu_{\alpha}(0)\rangle=|\nu_{\alpha}\rangle,
\end{equation}
where the Hamiltonian (after extracting constant diagonal matrix irrelevant for flavor 
transition as it generates a phase common to all flavors) 
is given as
\begin{widetext}
\begin{equation}
\mathcal{H} =  \frac{1}{2E} \left[ U \left( 
\begin{array}{ccc}
0 & 0 & 0 \\
0 &\Delta m_{21}^2 & 0 \\ 
0 & 0 & \Delta m_{31}^2
\end{array}
\right) U^{\dagger} + A \left( \begin{array}{ccc} 1 
& 0 &0 
\\ 0 & 0 & 
0 \\ 0 & 
0 & 0 \end{array} \right)
\right]. 
\label{Hamiltonian}
\end{equation}
\end{widetext}
The inclusion of the standard matter effect to the Hamiltonian is commonly known as the 
MSW effect\cite{Wolfenstein,Mikheyev}. Here 
\begin{widetext}
\begin{eqnarray}
U = U_{23} U_{13} U_{12} = 
\left[
\begin{array}{ccc}
1 & 0 &  0  \\
0 & c_{23} & s_{23} \\
0 & - s_{23} & c_{23} \\
\end{array}
\right] 
\left[
\begin{array}{ccc}
c_{13}  & 0 &  s_{13} e^{- i \delta}   \\
0 & 1 & 0 \\
- s_{13} e^{ i \delta}  & 0 & c_{13}  \\
\end{array}
\right] 
\left[
\begin{array}{ccc}
c_{12} & s_{12}  &  0  \\
- s_{12} & c_{12} & 0 \\
0 & 0 & 1 \\
\end{array}
\right] ,
\label{MNSmatrix}
\end{eqnarray}
\end{widetext} 
is the PMNS\cite{Pontecorvo,Maki} matrix in vacuum. $A=2EV_{cc}$ represents the 
interaction of the neutrino with matter, more precisely with electrons. $E$ is the 
energy of the neutrino, $V_{cc}$ represents the charge current interaction and given 
by $V_{cc}=\sqrt{2}G_{F}N_{e}$, where $G_{F}$ is the Fermi coupling constant and 
$N_{e}$ is the electron number density. By taking $\Delta m^2_{31}$ outside the 
square brackets, from Eq. (\ref{Hamiltonian}), we redefine the matter interaction 
as $\hat A=A/\Delta m^2_{31}$ and define $\alpha=\frac{\Delta m^2_{21}}
{\Delta m^2_{31}}\sim\epsilon$. For the T2K experiment, 
$\hat A= 0.06\simeq \epsilon$. \\

From Eq.(\ref{Hamiltonian}), let us first consider the case where NSI is absent.
As a method to simplify calculations, it is convenient to work in the tilde basis, 
which we define as $\tilde\nu_{\alpha}=(U_{23}^{\dagger}) _{\alpha\beta}\nu_{\beta}$. 
In this basis the Hamiltonian, consisting only the standard matter interaction part, or 
$\mathcal {H}_{M}$ becomes,
\begin{equation}
\tilde {\mathcal {H}}_{M}=U_{23}^{\dagger} \mathcal {H}_{M} U_{23},
\end{equation}
where we have defined $U_{23}$ in Eq.(\ref{MNSmatrix}). 
This Hamiltonian in the tilde basis can now be written as a sum of the Hamiltonians 
of different orders $(\tilde {\mathcal {H}}_{M}=\tilde {H}_{0}+\tilde {H}_{1})$, where 
the ordering is done with respect to $\sqrt{\epsilon}$. For example the zeroth order 
Hamiltonian, as a function of standard matter interaction looks like,
\begin{eqnarray} 
\tilde{H}_{0} &=& \frac{\Delta m^2_{31}}{2E}
\left[
\begin{array}{ccc}
0 & 0 & 0 \\
0 & 0 & 0 \\
0 & 0 & 1
\end{array}
\right], 
\label{H0}
\end{eqnarray}
and similarly,
\begin{widetext}
\begin{eqnarray} 
\tilde{H}_{1} &=& 
\frac{\Delta m^2_{31}}{2E} 
\left[
\begin{array}{ccc}
0 & 0 & s_{13} e^{ -i \delta} \\
0 & 0 & 0 \\
s_{13} e^{ i \delta} & 0 & 0 
\end{array}
\right] 
+
\frac{\Delta m^2_{31}}{2E}   \left[
\begin{array}{ccc}
\hat A+ \alpha s^2_{12} + s^2_{13} & \alpha  c_{12} s_{12}  & 0 \\
\alpha  c_{12} s_{12}  & \alpha  c^2_{12}  & 0 \\
0 & 0 & - s^2_{13}  
\end{array}
\right] 
\nonumber \\
&-&  
\frac{\Delta m^2_{31}}{2E} 
\left[
\begin{array}{ccc}
0  & 0 & \left( \alpha  s^2_{12} + \frac{1}{2} s^2_{13} \right) 
s_{13} e^{ -i \delta}  \\
0 & 0 & \alpha c_{12} s_{12} s_{13} e^{ -i \delta}   \\
\left( \alpha  s^2_{12} + \frac{1}{2} s^2_{13} \right) s_{13} 
e^{ i \delta}  & \alpha c_{12} s_{12} s_{13} e^{ i \delta}  & 0  
\end{array}
\right] 
\nonumber \\ 
&-&
\frac{\Delta m^2_{31}}{2E} \alpha  \left[
\begin{array}{ccc}
 s^2_{12} s^2_{13} & \frac{1}{2} c_{12} s_{12} s^2_{13} & 0 \\
 \frac{1}{2} c_{12} s_{12} s^2_{13} & 0  & 0 \\
0 & 0 & - s^2_{12} s^2_{13}
\end{array}
\right]. 
\label{H1}
\end{eqnarray}
\end{widetext}
Here $\tilde{H}_{1}$ is the perturbed part of the Hamiltonian in standard 
matter. The different matrices in the perturbed Hamiltonian in the tilde basis comprises 
of four different orders in 
$\sqrt{\epsilon}$, which are $\sqrt{\epsilon}$, $\epsilon$, $\epsilon^{\frac{3}{2}}$, 
and $\epsilon^2$ respectively.\\

Now we include the NSI matrix during propagation. 
The Hamiltonian consisting of these NSI parameters has the form\cite{Meloni},
\begin{eqnarray} 
\mathcal {H}_{NSI} = \frac{\Delta m^2_{31}}{2E}\hat A  
\left[
\begin{array}{ccc}
\epsilon^m_{e e} & \epsilon^m_{e \mu} & \epsilon^m_{e \tau} \\
\epsilon^{m*}_{e \mu} & \epsilon^m_{\mu \mu} & \epsilon^m_{\mu \tau} \\
\epsilon^{m*}_{e \tau} & \epsilon^{m*}_{\mu \tau} & \epsilon^m_{\tau \tau} 
\end{array}
\right]. 
\label{H-NSI}
\end{eqnarray}
where,
\begin{equation}
\epsilon^m_{\alpha \beta} = \sum_{f,P}\epsilon^{f P}_{\alpha \beta}\frac{n_f}{n_e}
\end{equation}
where $n_f$ is the number density of the fermion $f$ \cite{genbound-nsi}.  
Here $\epsilon^m_{\alpha \beta}$ ,
$(\alpha,\beta=e, \mu, \tau)$ are non-standard interaction (NSI) parameters of neutrinos, 
propagating through matter, defined as $\epsilon^m_{\alpha\beta}=|\epsilon^m_{\alpha\beta}
|e^{i\phi_{\alpha\beta}}$.
To include the $\mathcal {H}_{NSI}$ matrix in the perturbative part of the Hamiltonian, 
we have to first rotate $\mathcal {H}_{NSI}$ matrix from its flavor basis to the 
tilde basis by,
\begin{eqnarray}
\tilde{\mathcal {H}}_{NSI}=U^{\dagger}_{23}\mathcal {H}_{NSI}U_{23}.
\end{eqnarray}

Thus our total Hamiltonian $(\tilde{\mathcal {H}}=\tilde{\mathcal{H}}_{M}
+\tilde{\mathcal{H}}_{NSI})$, in the tilde basis can be written as a linear superposition 
of the zeroth order Hamiltonian($\tilde{H}_{0}$) with its perturbative parts in that 
same basis. After the inclusion of the NSI matrix, which has its effects at the 
sub leading part, we now redefine our Hamiltonian in the perturbative limit as 
($\tilde {H}_{1}\rightarrow\tilde {H}_{1}+\tilde{\mathcal {H}}_{NSI}$).
Since the upper bounds of these NSI parameters are quite high\cite{genbound-nsi}, 
therefore we will consider two different cases, one with $\epsilon_{\alpha\beta}^m\sim\sqrt
{\epsilon}$ and the other with $\epsilon_{\alpha\beta}^m\sim\epsilon$. 
Once we write the Hamiltonian in the tilde basis in this form, we would then look to 
evaluate the S-matrix. The S matrix in the tilde basis is related to the S matrix in the 
flavor basis by, $S(L)=U_{23}\tilde 
S(L) U_{23}^{\dagger},$ where $\tilde S(L)=T \text{exp} \left[ -i \int^{L}_{0} dx 
\tilde{\mathcal {H}} (x)  \right]$ and $L$ is the distance traversed. To evaluate $\tilde S(L)$ 
perturbatively, we 
choose $\Omega(x)$ as $\Omega(x) = e^{i \tilde{H}_{0} x} \tilde{S} (x)$, where $\Omega(x)$ 
obeys the evolution equation,
\begin{eqnarray} 
i \frac{d}{dx} \Omega(x) = H_{1} \Omega(x). 
\label{omega-evolution}
\end{eqnarray}
and $H_{1}$ is written in the form,
\begin{eqnarray} 
H_{1} \equiv e^{i \tilde{H}_{0} x} \tilde{H}_{1} e^{-i \tilde{H}_{0} x}. 
\label{def-H1}
\end{eqnarray}
From (\ref {omega-evolution}), we would like to deduce $\Omega(x)$ perturbatively. 
So the solution of the evolution equation followed by $\Omega(x)$, can be written
in terms of the $H_{1}$ matrices as,
\begin{widetext}
\begin{eqnarray} 
\Omega(x) &=& 1 + 
(-i) \int^{x}_{0} dx' H_{1} (x') + 
(-i)^2 \int^{x}_{0} dx' H_{1} (x') \int^{x'}_{0} dx'' H_{1} (x'') 
\nonumber \\
&+& (-i)^3 \int^{x}_{0} dx' H_{1} (x') \int^{x'}_{0} dx'' H_{1} (x'') 
\int^{x''}_{0} dx''' H_{1} (x''') + 
\mathcal{O} ( \epsilon^4 ). 
\label{Omega-exp}
\end{eqnarray}
\end{widetext}
From our previous definition of $\Omega(x)$, we can now write 
the S-matrix as,
\begin{eqnarray} 
\tilde{ S }(x) = e^{- i \tilde{H}_{0} x} \Omega(x). 
\label{Smatrix-tilde2}
\end{eqnarray}
The S-matrix in the flavor basis is obtained by rotating 
$\tilde S$ in the (2-3) space as $S=U_{23}\tilde S U_{23}^{\dagger}$.\\

Since the S-matrix changes the flavor of a neutrino state after traversing a length 
$L$, which is given by the expression,
\begin{eqnarray}
\nu_{\alpha}(L)=S_{\alpha \beta}\nu_{\beta}(0),
\end{eqnarray}
the oscillation probability of the neutrino, changing the flavor from 
$\alpha\rightarrow\beta$ is given as,
\begin{eqnarray}
P(\nu_{\beta}\rightarrow\nu_{\alpha};L)=|S_{\alpha\beta}|^2.
\label{probability}
\end{eqnarray} \\

This expression of the oscillation probability takes into consideration the standard 
matter interaction and the NSI during propagation only. In section III we will 
introduce the idea of NSI at the source and detector. It should be noted that since 
$\theta_{12}$ and $\theta_{23}$ are quite large, compared to $\theta_{13}$, they are 
considered to be in the zeroth order. 
\section{  NSI at source, detector and during propagation in 
$\nu_{\mu}\rightarrow \nu_{e}$ oscillation probability}
In the presence of NSI at the source and at the detector, the neutrino states produced 
at the detector can be treated as a superposition \cite{Zhang,Kopp-thesis} of pure 
orthonormal flavor states.
\begin{eqnarray}\label{eq:normalization}
|\nu^s_\alpha \rangle & = & \frac{1}{{N^s_\alpha}} \left( |\nu_\alpha
\rangle + \sum_{\beta=e,\mu,\tau} \epsilon^s_{\alpha\beta}
|\nu_\beta\rangle  \right) \ , \\ \langle \nu^d_\beta| & = &
\frac{1}{{N^d_\beta}} \left( \langle
 \nu_\beta | + \sum_{\alpha=e,\mu,\tau}
\epsilon^d_{\alpha \beta} \langle  \nu_\alpha  | \right) \ ,
\end{eqnarray}
where, $\epsilon^s_{\alpha\beta}$ and $\epsilon^d_{\alpha \beta}$ are NSI at source 
and detector, respectively and the normalization factors are given by
\begin{eqnarray}
\label{eq:factor}
{N^{s}_\alpha} & = & \sqrt{\left[\left(   \mathbf{1} + \epsilon^{s}
\right)\left(   \mathbf{1} + \epsilon^{s \dagger} \right)
\right]_{\alpha\alpha}} \nonumber , \\
{N^{d}_\beta} & = & \sqrt{\left[\left(   \mathbf{1} + \epsilon^{d
\dagger} \right)\left(   \mathbf{1} + {\epsilon^{d}} \right)
\right]_{\beta \beta}}   \ .
\end{eqnarray}
For example, $\epsilon_{\alpha\beta}^s$ describes a non-standard admixture of flavor 
$\beta$ to the neutrino state which is produced in association with a charged lepton 
of flavor $\alpha$. This means the neutrino source does not produce a pure flavor 
neutrino eigenstate $|\nu_{\alpha}\rangle$, but rather a superposition of pure 
orthonormal flavor states \cite{Kopp}. To be consistent with the literature the 
convention that we have chosen is in $\epsilon_{\alpha\beta}^s$, the first index 
corresponds to the flavor of the charged lepton, and the second to that of the neutrino, 
while in $\epsilon_{\alpha\beta}^d$, the order is reversed. In general, as we can 
clearly see from the definitions above that the matrices $(\mathbf{1}+\epsilon^s)$ 
and $(\mathbf{1}+\epsilon^d)$ are non unitary, i.e. the source and the detection states 
do not require to form a complete orthonormal sets of basis vectors in the Hilbert space. \\

Since the coefficients $\epsilon_{e\alpha}^s$ and $\epsilon_{\alpha e}^d$ both originate 
from the $(V-A)(V\pm A)$ coupling\cite{Kopp} to up and down quarks, so we have the 
constraint 
\begin{equation}
\epsilon_{e\alpha}^s=\epsilon_{\alpha e}^{d *}.
\end{equation}
Thus this condition reduces the number of independent parameters and makes the model 
more predictive. But in our paper we have presented the most general case.\\ 

Considering the NSI effects at the source, detector as well as during the propagation of
neutrinos through matter the amplitude of the oscillation becomes 
\begin{eqnarray}
P_{\nu^s_\alpha \rightarrow \nu^d_\beta}
&=& |\langle {\nu^d_\beta}|S(L)|{\nu^s_\alpha}\rangle|^2 \nonumber\\
&=& |\frac{1}{N^s_\alpha N^d_\beta} (1 + \epsilon^d)_{\gamma\beta} \, (S(L))_{\gamma\delta}
(1 + \epsilon^s)_{\alpha\delta} |^2           \nonumber\\
&=& |\frac{1}{N^s_\alpha N^d_\beta} [ ( 1 + \epsilon^d )^T \,\, S(L) \,\,
( 1 + \epsilon^s )^T ]_{\beta\alpha} |^2,
\label{eq:P-ansatz}
\end{eqnarray}
where the  $S(L)$ is defined earlier. Considering 
NSI at the source and at the detector of the order of $\epsilon$\cite{genbound-nsi}, 
we can now write the probability expression as a sum of the probabilities of different 
order $\sqrt{\epsilon}$ terms. The total oscillation probability would look like 
\begin{eqnarray}
P (\nu_\alpha \rightarrow \nu_\beta) &=& 
P_{\alpha \beta}^{(0)} + P_{\alpha \beta}^{(1/2)} + P_{\alpha \beta}^{(1)} + 
P_{\alpha \beta}^{(3/2)} + P_{\alpha \beta}^{(2)} . \nonumber \\ 
\label{Palpha-beta-def}
\end{eqnarray}

In our later analysis we would incorporate the results from the reactor neutrino experiments along 
with the long baseline superbeam experiment such as T2K. It is notable that the 
non-standard interaction parameters during propagation do not play any substantial 
role in case of the reactor neutrino experiments, due to its very short baseline. 
Furthermore, the NSI parameters present both at the source and at the detector of these 
two different kinds of neutrino experiments
i.e. $\epsilon^s_{\alpha\beta}$ and 
$\epsilon^d_{\alpha\beta}$ are considered to be the same.
The oscillation probability $P(\nu_{\alpha}
\rightarrow \nu_{\beta})$ is for a neutrino, rather than an antineutrino. However, 
one can relate the oscillation  probabilities for antineutrinos to those for neutrinos 
by 
\begin{equation}
P_{\bar{\alpha}\bar{\beta}}= P_{\alpha \beta}(\delta_{CP} \rightarrow 
-\delta_{CP}, {\hat A} \rightarrow 
- {\hat A}).
\label{probsd}
\end{equation}
In addition, we also have to replace $\epsilon^s$, $\epsilon^d$, $\epsilon_{\alpha\beta}^m$, 
with their
complex conjugates, in order to deduce the oscillation probability for the antineutrino, 
if one considers non-standard interaction during propagation and at the source and 
detector of the experiment. \\

It should be noted that the expression (\ref{eq:P-ansatz}) is also valid in the Minimal 
Unitarity Violation (MUV) model and is very instructive for analyzing the {\it CP} 
violating effects in the MUV model in future long baseline experiments
\cite{antus,Toshev,Goswami,Xing,Luo,Altarelli,Yasuda}. \\
\section{Perturbation theory by considering large NSI parameters during 
propagation}
In our next two sections, we will consider two different cases of these NSI parameters 
and present oscillation probability for the channel $\nu_{\mu}\rightarrow
\nu_{e}$, as was observed by the T2K experiment. 
In this section, we will consider $\epsilon_{\alpha\beta}^m\sim \sqrt{\epsilon}\sim 0.18$, and in 
the next section, we will put $\epsilon_{\alpha\beta}^m\sim\epsilon\sim 0.03$. \\

Since $\hat A$ is of the order of $\epsilon$, thus $\mathcal {H}_{NSI}$ is in the 
perturbative range $\epsilon^\frac{3}{2}$. Furthermore, we also have to transform 
$\mathcal {H}_{NSI}$ from its flavor basis to the tilde basis, e.g. 
\begin{eqnarray}
\tilde{\mathcal{H}}_{NSI}=U_{23}^{\dagger} \mathcal{H}_{NSI} U_{23}.
\end{eqnarray}
Our total Hamiltonian now looks like 
\begin{eqnarray}
\tilde{\mathcal {H}}=\tilde{H}_{0}+\left[\tilde{H}_{1}+
\tilde{\mathcal {H}}_{NSI}\right].
\end{eqnarray}
We include this $\tilde{\mathcal{H}}_{NSI}$ in the perturbative part of the Hamiltonian
and follow the same calculations described previously. The order $\epsilon^{3/2}$ 
component of the Hamiltonian now looks like
\begin{eqnarray}
&&\tilde{H_{1}}(\epsilon^{3/2})=
-\frac{\Delta m^2_{31}}{2E} s_{13}\nonumber \\
&&\left[
\begin{array}{ccc}
0  & 0 & (\alpha s^2_{12}+\frac{1}{2}s^2_{13})e^{-i\delta}  \\
0 & 0 & \alpha c_{12}s_{12}e^{-i\delta}   \\
(\alpha s^2_{12}+\frac{1}{2}s^2_{13})e^{i\delta}  & \alpha c_{12}s_{12}e^{i\delta}  & 0  
\end{array}
\right] 
\nonumber \\
&&+\frac{\Delta m^2_{31}}{2E} 
\hat A
U_{23}^{\dagger}
\left[
\begin{array}{ccc}
\epsilon^m_{e e} & \epsilon^m_{e \mu} & \epsilon^m_{e \tau} \\
\epsilon^{m*}_{e \mu} & \epsilon^m_{\mu \mu} & \epsilon^m_{\mu \tau} \\
\epsilon^{m*}_{e \tau} & \epsilon^{m*}_{\mu \tau} & \epsilon^m_{\tau \tau} 
\end{array}
\right]
U_{23}.
\end{eqnarray}
By computing the S-matrix, 
which comprises of the standard matter interaction, and NSI during propagation of 
the neutrino, we then include the NSI parameters at the source and detector, which 
are of the order of $\epsilon$, as per the present bounds on these parameters suggest.
Finally we write down the oscillation probability for the muon neutrino going to 
electron neutrino up-to second order in $\epsilon$. It is noteworthy that due to the
non unitarity of the non-standard interaction matrices at the source and at the detector,
the Probability of neutrino oscillation is not normalized to unity. So one has to 
include necessary normalization factors as we have done in (\ref {eq:factor}). In the 
context of T2K, where we are observing muon neutrino oscillation to electron neutrino,
these normalization factors do not play a very significant role. To be precise 
the effects of these normalization terms are greater than $\mathcal {O}
(\epsilon^2)$, which we are neglecting. Finally by considering all these effects,
the oscillation probability in the $\nu_{\mu}\rightarrow\nu_{e}$ channel is,
\begin{widetext}
\begin{align}
\label{prob_large_epsilon}
P_{\nu_{\mu}\rightarrow \nu_e}&=|\epsilon^d_{e \mu}|^2+|\epsilon^s_{e \mu}|^2+
2 |\epsilon^d_{e \mu}| |\epsilon^s_{e \mu}| \cos[\phi^d_{e \mu}-\phi^s_{e \mu}]+
\frac{L^2 \alpha^2 \Delta m^4_{31} c_{23}^2 s_{2 \times 12}^2}{16 E^2}\nonumber \\
&+\frac{L\alpha  \Delta m^2_{31} |\epsilon^d_{e \tau}| c_{23}^2}{E} 
\cos\left[\frac{L \Delta m^2_{31}}{4 E}+\phi^d_{e \tau}\right] 
\sin\left[\frac{L \Delta m^2_{31}}{4 E}\right] s_{2 \times 12} s_{23}\nonumber \\
&-2|\epsilon^d_{e \mu}| |\epsilon^s_{e \mu}| \cos[\phi^d_{e \mu}] 
\cos[\phi^s_{e \mu}] s_{23}^2+2 |\epsilon^d_{e \mu}| |\epsilon^s_{e \mu}| 
\cos\left[\frac{L \Delta m^2_{31}}{2 E}\right] \cos[\phi^d_{e \mu}] \cos[\phi^s_{e \mu}] 
s_{23}^2\nonumber \\
&+8 a_3 \cos[\delta +\phi_{a_3}] \sin^2\left[\frac{L
\Delta m^2_{31}}{4 E}\right] s_{13} s_{23}^2+8 |\epsilon^d_{ee}| \sin^2\left[\frac{L
\Delta m^2_{31}}{4 E}\right] s_{13}^2 s_{23}^2+8 |\epsilon^s_{\mu \mu}| \sin^2\left[\frac{L
\Delta m^2_{31}}{4 E}\right] s_{13}^2 s_{23}^2\nonumber \\
&+\frac{ s_{13}^2 s_{23}^2}{E}\sin\left[\frac{L \Delta m^2_{31}}{4 E}\right] 
\left(-2 A L \Delta m^2_{31} \cos\left[\frac{L \Delta m^2_{31}}{4 E}\right]+
2 E (1+4 A+c_{2\times 13}) \sin\left[\frac{L \Delta m^2_{31}}{4 E}\right]\right)\nonumber \\
&+4 |\epsilon^d_{e \tau}| c_{23} \sin^2\left[\frac{L \Delta m^2_{31}}{4 E}\right]
(|\epsilon^d_{e \tau}| c_{23}+2 \cos[\delta -\phi^d_{e \tau}] s_{13}) s_{23}^2\nonumber \\
&+\frac{L \alpha  \Delta m^2_{31} s_{13} s_{23}}{E} \left(\cos\left[\delta +
\frac{L \Delta m^2_{31}}{4 E}\right] c_{23} \sin\left[\frac{L \Delta m^2_{31}}{4 E}\right] 
s_{2\times 12}-\sin\left[\frac{L \Delta m^2_{31}}{2 E}\right] s_{12}^2 s_{13} 
s_{23}\right)\nonumber \\
&+\frac{a_2 L \Delta m^2_{31}}{E} \cos\left[\delta +\frac{L \Delta m^2_{31}}{4 E}+
\phi_{a_2}\right]
\sin\left[\frac{L \Delta m^2_{31}}{4 E}\right] s_{13} s_{2\times 23}\nonumber \\
&-2 |\epsilon^d_{e \mu}| |\epsilon^d_{e \tau}| \sin\left[\frac{L \Delta m^2_{31}}{4 E}\right] 
s_{2\times 23} \left(c_{2\times 23}  \cos[\phi^d_{e \tau}-\phi^d_{e \mu}] 
\sin\left[\frac{L \Delta m^2_{31}}{4 E}\right]+\cos\left[\frac{L
\Delta m^2_{31}}{4 E}\right] \sin[\phi^d_{e \tau}-\phi^d_{e \mu}]\right)\nonumber \\
&-2 |\epsilon^d_{e \mu}| |\epsilon^s_{e \mu}| \cos[\phi^s_{e \mu}] 
\sin\left[\frac{L \Delta m^2_{31}}{2 E}\right] s_{23}^2 \sin[\phi^d_{e \mu}]\nonumber \\
&-2 s_{23} \left(|\epsilon^d_{e \mu}| \cos[\phi^d_{e \mu}] \sin[\delta ] \sin\left[\frac{L
\Delta m^2_{31}}{2 E}\right] s_{13}+|\epsilon^d_{e \mu}| \cos[\delta ] s_{13}
\left(2 c_{2\times 23}  \cos[\phi^d_{e \mu}] \sin^2\left[\frac{L \Delta m^2_{31}}
{4 E}\right]\right.\right. \nonumber \\
&-\left.\left.\sin\left[\frac{L\Delta m^2_{31}}{2 E}\right] \sin[\phi^d_{e \mu}]
\right)+2 \sin^2\left[\frac{L \Delta m^2_{31}}{4 E}\right]
\left(-2 |\epsilon^s_{\mu \tau}| c_{23} \cos[\phi^s_{\mu \tau}] s_{13}^2+
|\epsilon^d_{e \mu}|^2 c_{23}^2 s_{23} \right.\right. \nonumber \\
&+\left.\left.|\epsilon^d_{e \mu}| c_{2\times 23}  \sin[\delta
] s_{13} \sin[\phi^d_{e \mu}]\right)\right)+\frac{L \alpha  \Delta m^2_{31} 
|\epsilon^d_{e \mu}| c_{23} s_{2 \times 12} }{2 E}\left(c_{23}^2 \sin[\phi^d_{e \mu}]+s_{23}^2
\sin\left[\frac{L \Delta m^2_{31}}{2 E}+\phi^d_{e \mu}\right]\right)\nonumber \\
&-4 |\epsilon^s_{e \mu}| \sin\left[\frac{L
\Delta m^2_{31}}{4 E}\right] s_{13} s_{23} \sin\left[\delta +\frac{L \Delta m^2_{31}}{4 E}-
\phi^s_{e \mu}\right]\nonumber \\
&-2 |\epsilon^d_{e \tau}| |\epsilon^s_{e \mu}| \sin\left[\frac{L \Delta m^2_{31}}{4
E}\right] s_{2\times 23} \sin\left[\frac{L \Delta m^2_{31}}{4 E}+\phi^d_{e \tau}-
\phi^s_{e \mu}\right]+\frac{L \alpha  \Delta m^2_{31} |\epsilon^s_{e \mu}| 
c_{12} c_{23} s_{12}] \sin[\phi^s_{e \mu}]}{E}\nonumber \\
&+2 |\epsilon^d_{e \mu}| |\epsilon^s_{e \mu}| \cos[\phi^d_{e \mu}] \sin\left[\frac{L
\Delta m^2_{31}}{2 E}\right] s_{23}^2 \sin[\phi^s_{e \mu}]-2 |\epsilon^d_{e \mu}| 
|\epsilon^s_{e \mu}| s_{23}^2 \sin[\phi^d_{e \mu}] \sin[\phi^s_{e \mu}]\nonumber \\
&+2 |\epsilon^d_{e \mu}| |\epsilon^s_{e \mu}| \cos\left[\frac{L \Delta m^2_{31}}{2 E}\right] 
s_{23}^2 \sin[\phi^d_{e \mu}] \sin[\phi^s_{e \mu}]
\end{align}
\end{widetext}
where $s_{2\times ij}=\sin 2\theta_{ij}$ and $c_{2\times ij}=\cos 2\theta_{ij}$ and

\begin{widetext}
\begin{align}
a_2 &=\frac{A}{\sqrt{2}}\sqrt{|\epsilon^m_{e \mu}|^2+|\epsilon^m_{e \tau}|^2+
\left(|\epsilon^m_{e \mu}|^2-|\epsilon^m_{e \tau}|^2\right)c_{2 \times 23}-
2 |\epsilon^m_{e \mu}| |\epsilon^m_{e \tau}| \cos[\phi^m_{e \mu}-\phi^m_{e \tau}] 
s_{2\times 23}} \;\; ,\nonumber \\
a_3 &=\frac{A}{\sqrt{2}} \sqrt{|\epsilon^m_{e \mu}|^2+|\epsilon^m_{e \tau}|^2+
\left(-|\epsilon^m_{e \mu}|^2+|\epsilon^m_{e \tau}|^2\right)
c_{2 \times 23}+2 |\epsilon^m_{e \mu}| |\epsilon^m_{e \tau}| \cos[\phi^m_{e \mu}-
\phi^m_{e \tau}] s_{2\times 23}}\;\; ,\nonumber \\
\phi_{a_2} &=\tan^{-1}\left[\frac{\epsilon^m_{e \mu} c_{23} \sin[\phi^m_{e \mu}]-
\epsilon^m_{e \tau} s_{23} \sin[\phi^m_{e \tau}]}{\epsilon^m_{e \mu} c_{23} 
\cos[\phi^m_{e \mu}]-\epsilon^m_{e \tau} \cos[\phi^m_{e \tau}] s_{23}}\right]\;\; ,\nonumber  \\
\phi_{a_3}&=\tan^{-1}\left[\frac{\epsilon^m_{e \mu} s_{23}\sin[\phi^m_{e \mu}]+
\epsilon^m_{e \tau} c_{23}\sin[\phi^m_{e \tau}]}{\epsilon^m_{e \tau} c_{23}
\cos[\phi^m_{e \tau}]+\epsilon^m_{e \mu}\cos[\phi^m_{e \mu}] s_{23}}\right]\; .
\end{align}
\end{widetext}
There are a few salient features of this expression of the probability.
These are as follows
\begin{itemize}
\item Considering the baseline length to be zero, we are left with the term
\begin{eqnarray}
P^{ND}_{\nu_{\mu}\rightarrow\nu_{e}}=
|\epsilon^d_{e\mu}|^2+|\epsilon^s_{e\mu}|^2
+2|\epsilon^d_{e\mu}||\epsilon^s_{e\mu}|\cos(\phi^d_{e\mu} - \phi^s_{e\mu}).\nonumber \\
\label{ND}
\end{eqnarray}
This term is the manifestation of the non-unitarity of the source and detector matrices,
more commonly know as the zero distance effect.
\item Assuming the standard matter interaction and NSI during the propagation,
as well as at the source and at the detector to be absent, we can obtain the expression 
of the probability, representing the vacuum oscillation
probability for a three flavor neutrino scenario correct upto  $\mathcal{O}(\alpha^2)$ . 
Particularly the following leading term of vacuum oscillation which is \cite{Gavela} 
\begin{eqnarray}
P^{Vacuum}_{\nu_{\mu}\rightarrow\nu_{e}}=
s^2_{2\times 13}s^2_{23}\sin^2\left
[\frac{\Delta m^2_{31}L}{4E}\right].
\label{Vacuum}
\end{eqnarray} 
can be obtained from eleventh term of \eqref{prob_large_epsilon} after considering $A \rightarrow 0$. 
\item 
Since muon is produced at the source therefore we observe only $\epsilon^s_{e\mu}$, 
$\epsilon^s_{\mu \mu}$, $\epsilon^s_{\mu\tau}$ NSI parameters, and electron is 
obtained at the detector, thus we observe $\epsilon^d_{e e}$, $\epsilon^d_{e\mu}$, 
$\epsilon^d_{e\tau}$ in our expression for the probability.
\item For the $\nu_{\mu}\rightarrow\nu_{e}$ channel, only $\epsilon^m_{e\mu}$ and 
$\epsilon^m_{e\tau}$ appears  NSI parameters during the propagation of the 
neutrino. The contribution from all the other NSI parameters during propagation 
are very much suppressed.
\item As mentioned, our expression of the probability is of the order of 
$\epsilon^2$, by considering large angle $\theta_{13}$. Similar expressions 
are to be found in\cite{Kopp}, the authors in there work considered small 
$\theta_{13}$. But the recent reactor based experiments\cite{Daya Bay,Reno}
compelled us to consider the regime of large $\sin\theta_{13}\sim\sqrt{\epsilon}$.
\end{itemize}
\section{Perturbation theory by considering small NSI parameters during propagation}
In this section we will concentrate on the idea of small non-standard interaction 
parameters during propagation. The standard matter interaction is again considered 
to be of the order of $\epsilon$, and we now consider NSI during propagation 
$\epsilon_{\alpha\beta}^m\sim\epsilon$. Following the same argument, as done in the previous section, 
the zeroth order Hamiltonian in the tilde basis remains the same. However the 
perturbative Hamiltonian gets rearranged. In this case $\mathcal {\tilde{H}}_{NSI}$ 
would be of the order of $\epsilon^2$. \\

The $\epsilon^2$ part of the perturbed Hamiltonian (\ref{H1}), can now be written as,
\begin{eqnarray}
\tilde {H}_{1}(\epsilon^2)=
&-&
\frac{\Delta m^2_{31}}{2E} \alpha  \left[
\begin{array}{ccc}
 s^2_{12} s^2_{13} & \frac{1}{2} c_{12} s_{12} s^2_{13} & 0 \\
 \frac{1}{2} c_{12} s_{12} s^2_{13} & 0  & 0 \\
0 & 0 & - s^2_{12} s^2_{13}
\end{array}
\right]\nonumber \\
&+&
\frac{\Delta m^2_{31}}{2E} 
\hat A
U_{23}^{\dagger}
\left[
\begin{array}{ccc}
\epsilon^m_{e e} & \epsilon^m_{e \mu} & \epsilon^m_{e \tau} \\
\epsilon^{m*}_{e \mu} & \epsilon^m_{\mu \mu} & \epsilon^m_{\mu \tau} \\
\epsilon^{m*}_{e \tau} & \epsilon^{m*}_{\mu \tau} & \epsilon^m_{\tau \tau} 
\end{array}
\right]U_{23}.
\end{eqnarray}
We again follow the same procedure as performed in the previous section. We 
computed the S-matrix, after including the standard matter interaction, and 
NSI during propagation. Then we considered the source and the detector effect.
Thus the probability for muon neutrino going to electron neutrino for 
$\epsilon^m_{\alpha\beta}\sim\epsilon$ is given by,
\begin{widetext}
\begin{align}
P_{\nu_{\mu}\rightarrow \nu_e}&=|\epsilon^d_{e \mu}|^2+|\epsilon^s_{e \mu}|^2+
2 |\epsilon^d_{e \mu}| |\epsilon^s_{e \mu}| \cos[\phi^d_{e \mu}-\phi^s_{e \mu}]+
\frac{L^2 \alpha ^2 \Delta m^4_{31} c_{23}^2 s_{2\times 12}^2}{16 E^2}\nonumber \\
&+\frac{L\alpha  \Delta m^2_{31} |\epsilon^d_{e \tau}| c_{23}^2 }{E}\cos
\left[\frac{L \Delta m^2_{31}}{4 E}+\phi^d_{e \tau}\right] \sin\left[\frac{L 
\Delta m^2_{31}}{4 E}\right] s_{2\times 12} s_{23}-2
|\epsilon^d_{e \mu}| |\epsilon^s_{e \mu}| \cos[\phi^d_{e \mu}] \cos[\phi^s_{e \mu}] 
s_{23}^2\nonumber \\
&+2 |\epsilon^d_{e \mu}| |\epsilon^s_{e \mu}| \cos\left[\frac{L \Delta m^2_{31}}{2 E}
\right] \cos[\phi^d_{e \mu}] \cos[\phi^s_{e \mu}] s_{23}^2+8 |\epsilon^d_{ee}| 
\sin^2\left[\frac{L \Delta m^2_{31}}{4E}\right] s_{13}^2 s_{23}^2\nonumber \\
&+8 |\epsilon^s_{\mu \mu}| \sin^2\left[\frac{L \Delta m^2_{31}}{4 E}\right] 
s_{13}^2 s_{23}^2 \nonumber \\
&+\frac{s_{13}^2 s_{23}^2}{E}\sin\left[\frac{L \Delta m^2_{31}}{4 E}\right]
\left(-2 A L \Delta m^2_{31} \cos\left[\frac{L \Delta m^2_{31}}{4 E}\right]+
2 E (1+4 A+c_{2 \times 13})
\sin\left[\frac{L \Delta m^2_{31}}{4 E}\right]\right)  \nonumber \\
&+4|\epsilon^d_{e \tau}| c_{23} \sin^2\left[\frac{L \Delta m^2_{31}}{4 E}\right] 
(|\epsilon^d_{e \tau}| c_{23}+2 \cos[\delta -\phi^d_{e \tau}] s_{13}) s_{23}^2\nonumber \\
&+\frac{L\alpha  \Delta m^2_{31} s_{13} s_{23} }{E}\left(\cos\left[\delta +
\frac{L \Delta m^2_{31}}{4E}\right] c_{23} \sin\left[\frac{L \Delta m^2_{31}}{4 E}\right] 
s_{2\times 12}-\sin\left[\frac{L\Delta m^2_{31}}{2 E}\right] s_{12}^2 s_{13} 
s_{23}\right) \nonumber \\ 
&-2|\epsilon^d_{e \mu}| |\epsilon^d_{e \tau}| \sin\left[\frac{L \Delta m^2_{31}}{4 E}\right] 
s_{2 \times 23} \left(c_{2 \times 23} \cos[\phi^d_{e \tau}-\phi^d_{e \mu}] 
\sin\left[\frac{L \Delta m^2_{31}}{4E}\right]+\cos\left[\frac{L \Delta m^2_{31}}{4 E}\right] 
\sin[\phi^d_{e \tau}-\phi^d_{e \mu}]\right) \nonumber \\
&-2|\epsilon^d_{e \mu}| |\epsilon^s_{e \mu}| \cos[\phi^s_{e \mu}] 
\sin\left[\frac{L \Delta m^2_{31}}{2 E}\right]
s_{23}^2 \sin[\phi^d_{e \mu}]-2 s_{23} \left(|\epsilon^d_{e \mu}| \cos[\phi^d_{e \mu}] 
\sin[\delta ] \sin\left[\frac{L \Delta m^2_{31}}{2 E}\right] s_{13} \right.\nonumber \\
&+\left. |\epsilon^d_{e \mu}| \cos[\delta ] s_{13} \left(2 c_{2 \times 23} 
\cos[\phi^d_{e \mu}] \sin^2\left[\frac{L
\Delta m^2_{31}}{4 E}\right]-\sin\left[\frac{L \Delta m^2_{31}}{2 E}\right] 
\sin[\phi^d_{e \mu}]\right) \right.\nonumber \\
&+\left.2\sin^2\left[\frac{L \Delta m^2_{31}}{4 E}\right] \left(-2 |\epsilon^s_{\mu \tau}| 
c_{23} \cos[\phi^s_{\mu \tau}] s_{13}^2+|\epsilon^d_{e \mu}|^2 c_{23}^2 s_{23}+
|\epsilon^d_{e \mu}| c_{2 \times 23} \sin[\delta ] s_{13} \sin[\phi^d_{e \mu}]
\right)\right)\nonumber \\
&+\frac{L\alpha  \Delta m^2_{31} |\epsilon^d_{e \mu}| c_{23} s_{2\times 12}}{2 E} 
\left(c_{23}^2 \sin[\phi^d_{e \mu}]+s_{23}^2 \sin\left[\frac{L \Delta m^2_{31}}{2 E}+
\phi^d_{e \mu}\right]\right)\nonumber \\
&-4 |\epsilon^s_{e \mu}| \sin\left[\frac{L \Delta m^2_{31}}{4 E}\right] 
s_{13} s_{23} \sin\left[\delta +\frac{L \Delta m^2_{31}}{4 E}-\phi^s_{e \mu}\right]\nonumber \\
&-2 |\epsilon^d_{e \tau}| |\epsilon^s_{e \mu}| \sin\left[\frac{L \Delta m^2_{31}}{4 E}\right] 
s_{2 \times 23} \sin\left[\frac{L
\Delta m^2_{31}}{4 E}+\phi^d_{e \tau}-\phi^s_{e \mu}\right]+\frac{L \alpha  \Delta m^2_{31} 
|\epsilon^s_{e \mu}|
c_{12} c_{23} s_{12} \sin[\phi^s_{e \mu}]}{E}\nonumber \\
&+2 |\epsilon^d_{e \mu}| |\epsilon^s_{e \mu}| \cos[\phi^d_{e \mu}] \sin\left[\frac{L 
\Delta m^2_{31}}{2 E}\right] s_{23}^2 \sin[\phi^s_{e \mu}]-2 |\epsilon^d_{e \mu}| 
|\epsilon^s_{e \mu}| s_{23}^2 \sin[\phi^d_{e \mu}] \sin[\phi^s_{e \mu}]\nonumber \\
&+2 |\epsilon^d_{e \mu}| |\epsilon^s_{e \mu}| \cos\left[\frac{L \Delta m^2_{31}}{2
E}\right] s_{23}^2 \sin[\phi^d_{e \mu}] \sin[\phi^s_{e \mu}].
\label{P2}
\end{align}

\end{widetext}
We would again try to emphasize on some of the interesting features about this 
oscillation probability expression. 
\begin{itemize}
\item As expected, we get back the same expression for the near detector 
effect, which was provided in the expression (\ref{ND}).
\item Similar to the result of previous section, assuming the standard matter 
interaction and NSI during the propagation,
as well as at the source and at the detector to be absent, we can obtain the the 
expression of the probability, representing the vacuum oscillation
probability for a three flavor neutrino scenario correct upto  $\mathcal{O}(\alpha^2)$ . 
Particularly from the tenth term in \eqref{P2} considering $A\rightarrow 0$ one can get 
the leading vacuum oscillation probability in \eqref{Vacuum}.
\item It is very much interesting to note, that due to the choice of the NSI 
parameters during propagation proportional to $\epsilon$, the probability expression 
up to second order in $\epsilon$ for this particular channel is devoid of any terms 
containing this kind of NSI. It shows that it is very difficult to constrain such small
NSIs in relatively short baseline neutrino oscillation experiments like T2K.
\item It is conspicuous, that the NSI parameters at the source and at the 
detector carry the same flavor indices, as in Eq.(\ref{prob_large_epsilon}).
\end{itemize}

\section{Numerical Analysis}


Here, we discuss the approach of our  complete numerical analysis in obtaining the 
results presented in this 
\begin{figure}[htb]
\includegraphics[width=6cm, height=4.5cm]{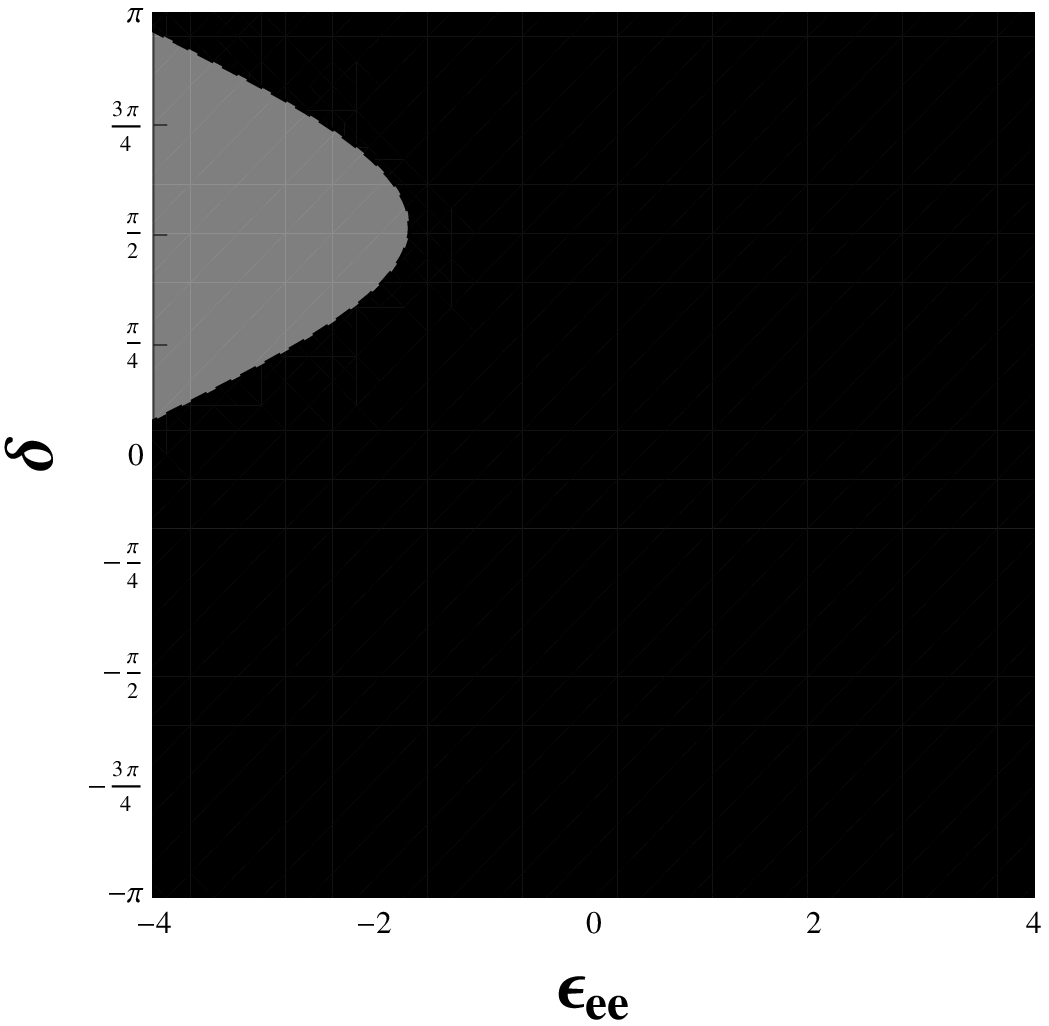}
\hspace{0.60cm} 
\includegraphics[width=6cm, height=4.5cm]{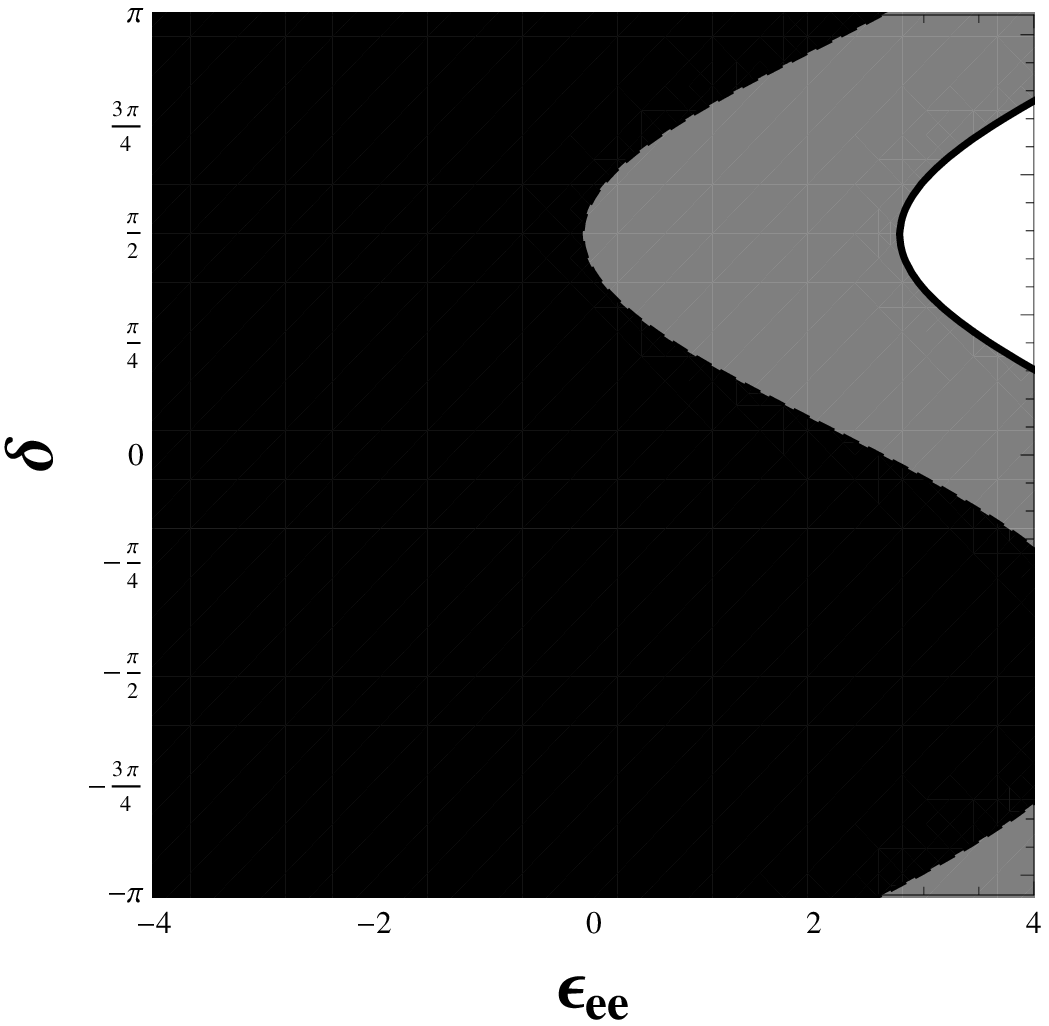}
\caption{Plot of $\epsilon_{ee} - \delta$ for  real NSI in matter. Excluded region 
(white at 90\% and Grey+white region at 66\% confidence level). Upper(lower) panel 
corresponds to normal(inverted) hierarchy.}
\label{fig:fig1}
\end{figure}
\begin{figure}[htb]
\includegraphics[width=6cm, height=4.5cm]{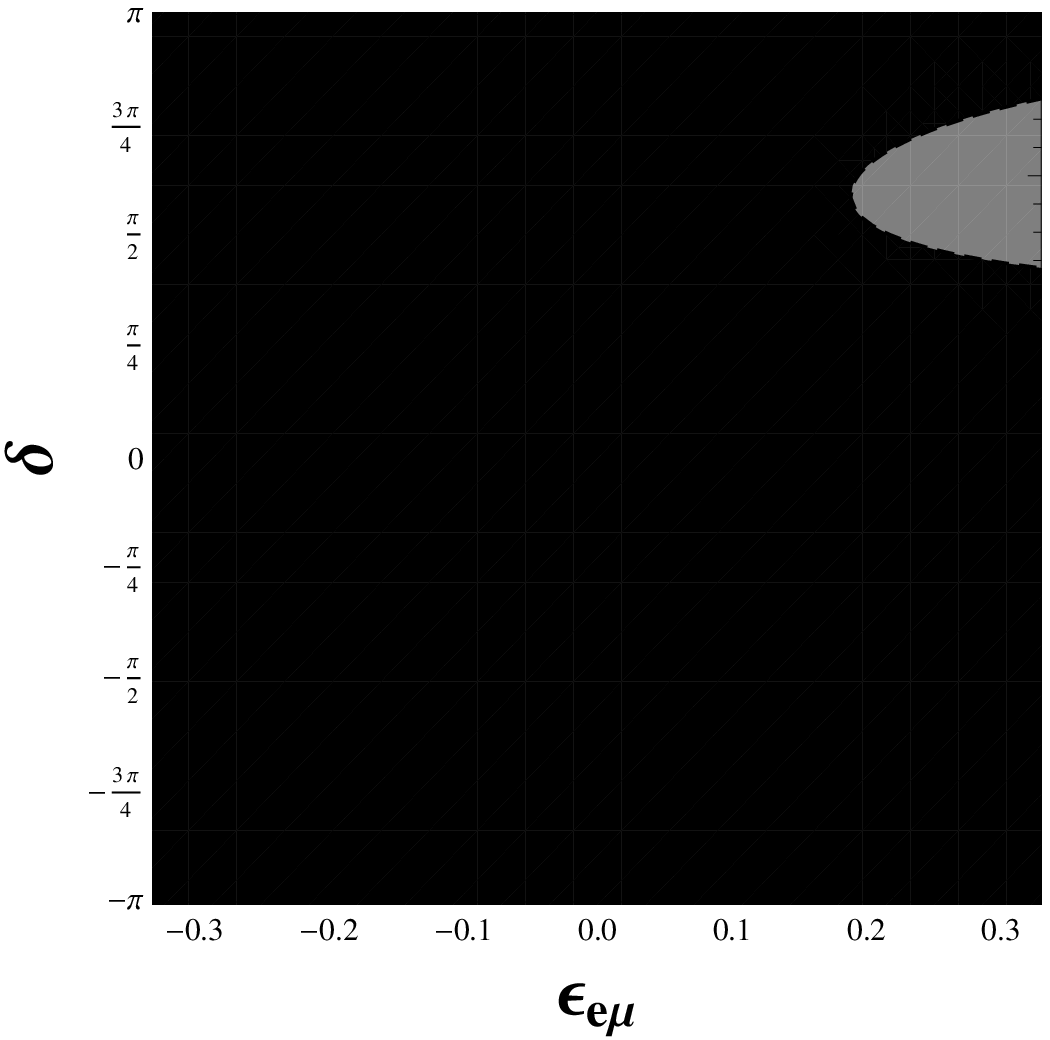}
\hspace{0.60cm}
\includegraphics[width=6cm, height=4.5cm]{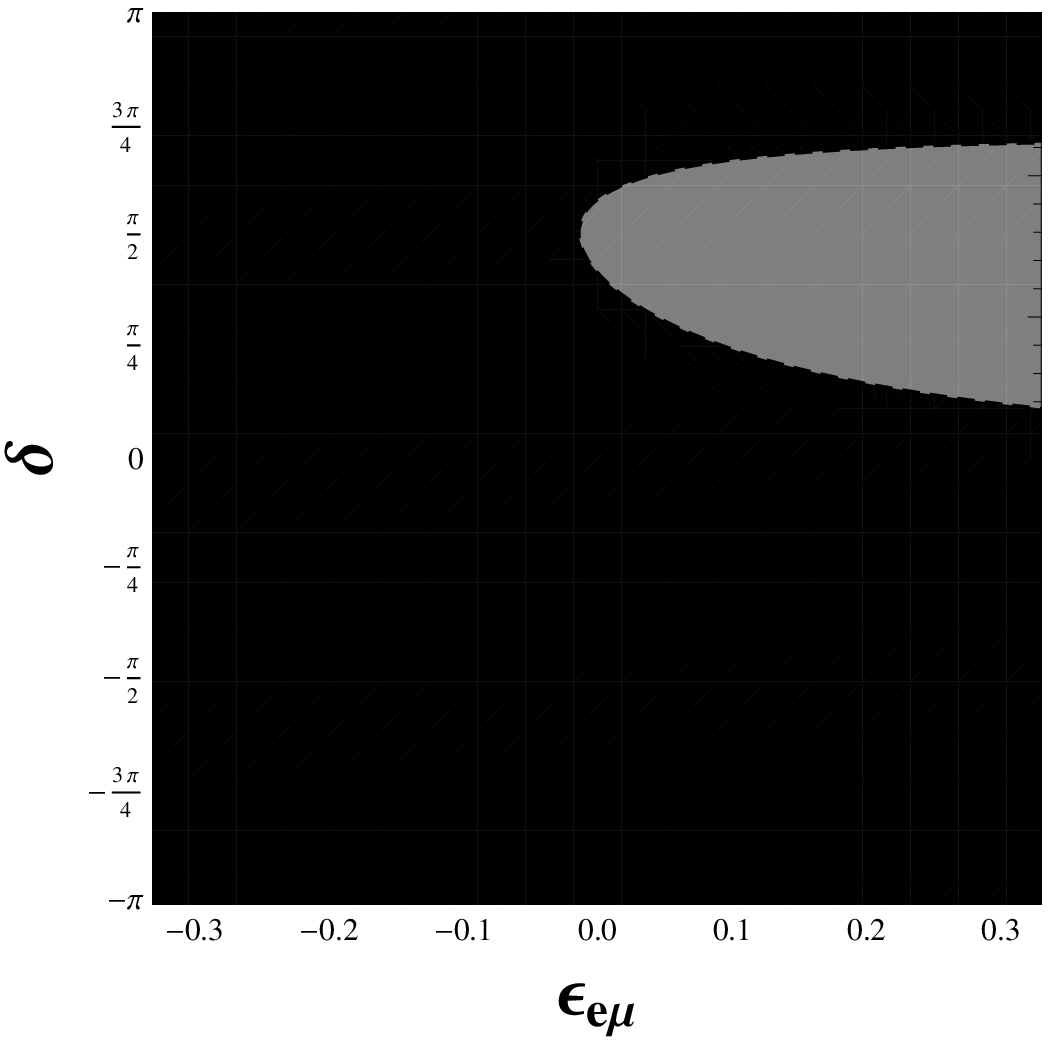}
\caption{Plot of $\epsilon_{e\mu} - \delta$ for  real NSI in matter. Excluded region 
(white at 90\% and Grey+white region at 66\% confidence level). Upper(lower) panel 
corresponds to normal(inverted) hierarchy.}
\label{fig:fig2}
\end{figure}
\begin{figure}[htb]
\includegraphics[width=6cm, height=4.5cm]{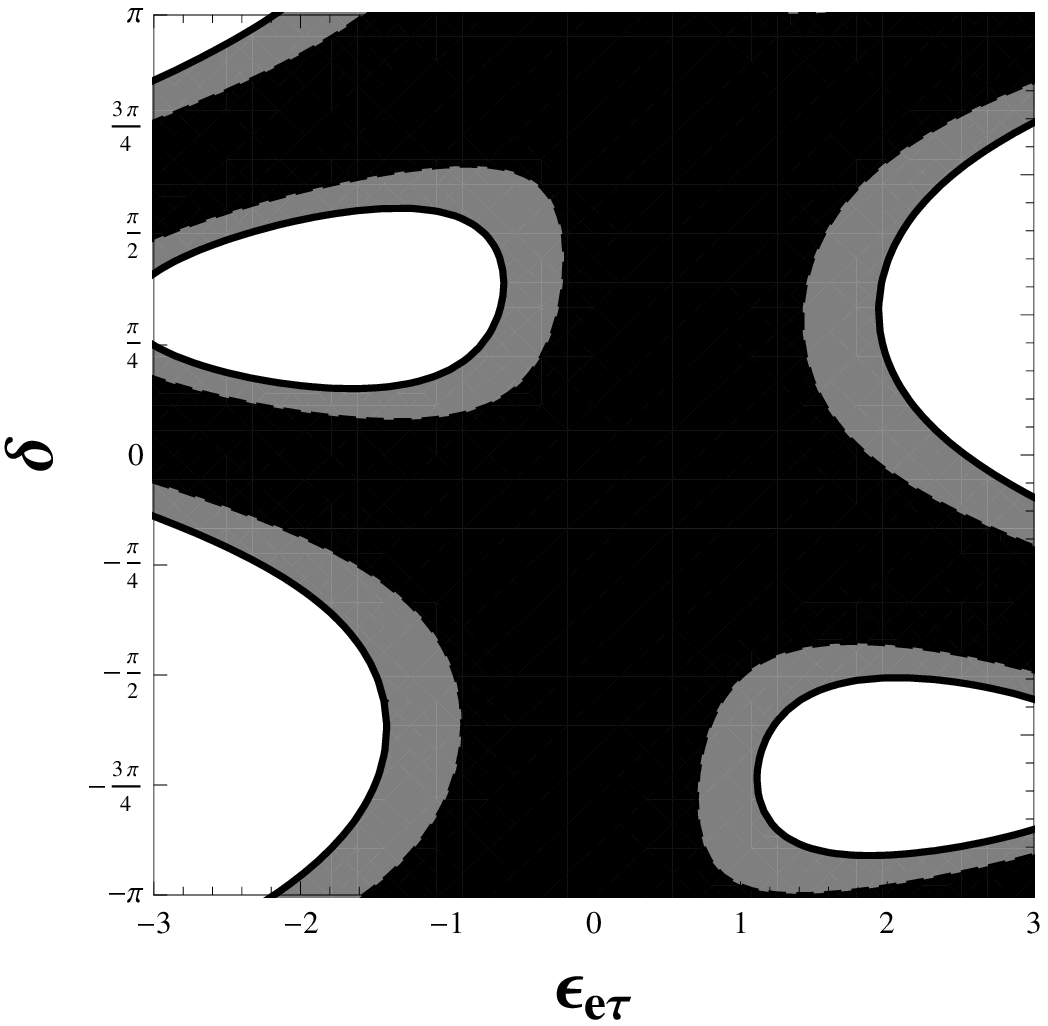}
\hspace{0.60cm}
\includegraphics[width=6cm, height=4.5cm]{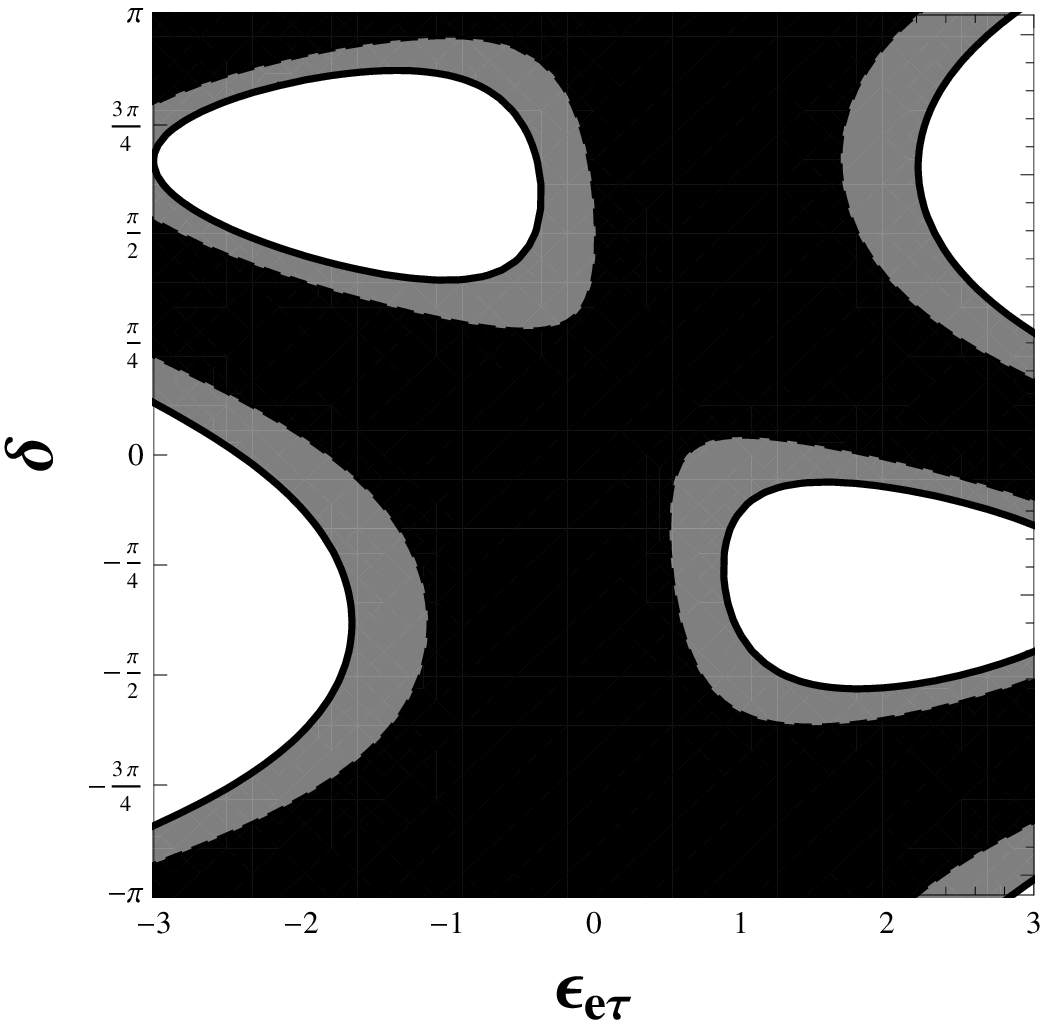}
\caption{Plot of $\epsilon_{e\tau} - \delta$ for  real NSI in matter. Excluded region 
(white at 90\% and Grey+white region at 66\% confidence level). Upper(lower) panel corresponds 
to normal(inverted) hierarchy.}
\label{fig:fig3}
\end{figure}
\begin{figure}[htb]
\includegraphics[width=6cm, height=4.5cm]{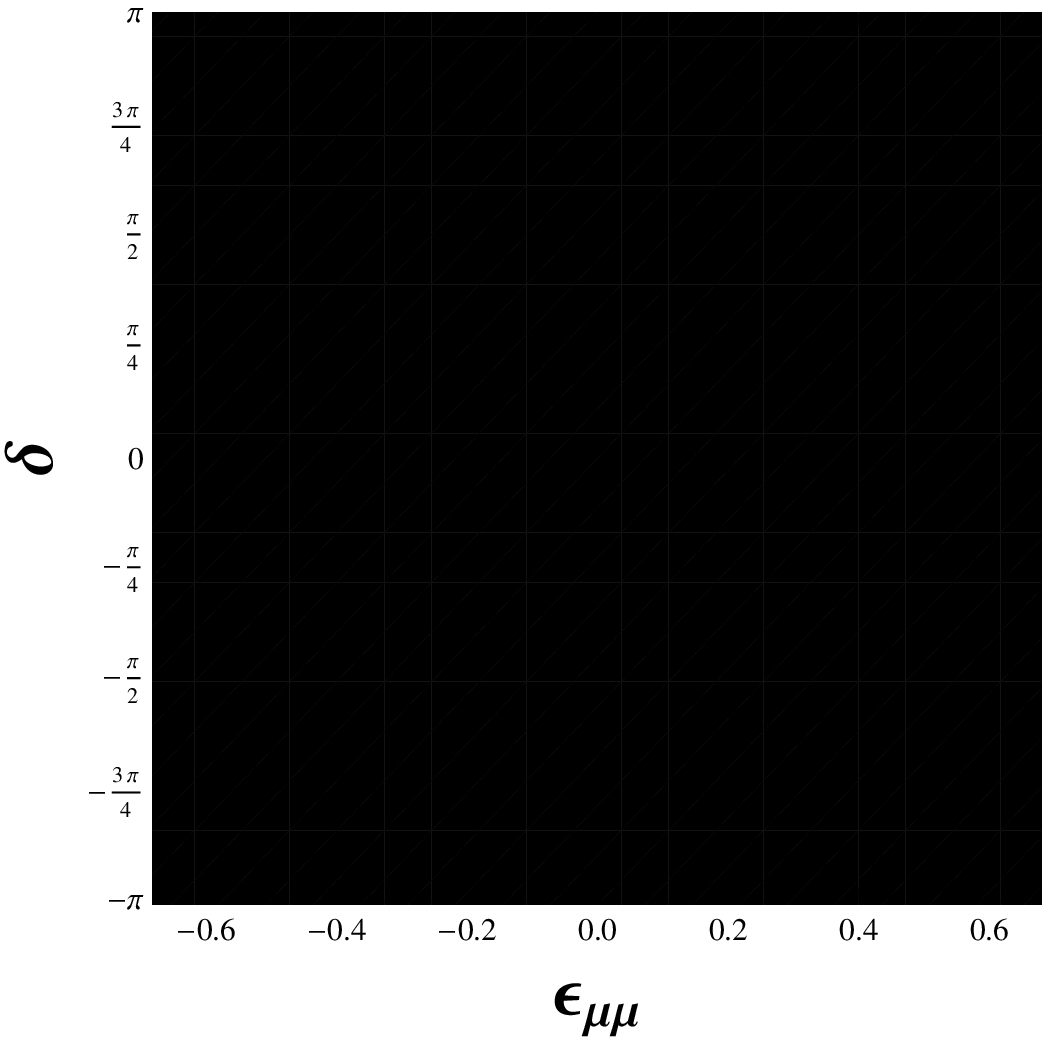}
\hspace{0.60cm}
\includegraphics[width=6cm, height=4.5cm]{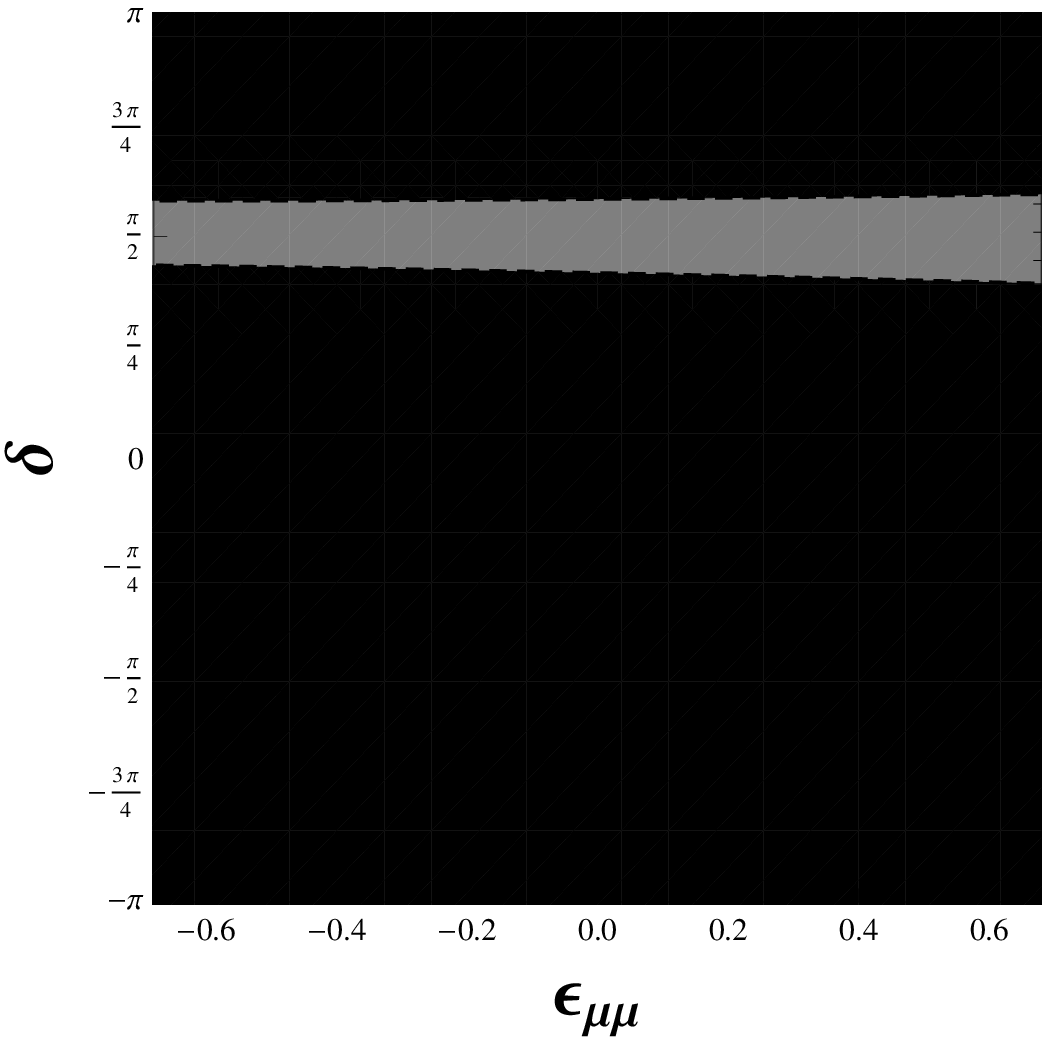}
\caption{Plot of $\epsilon_{\mu\mu} - \delta$ for  real NSI in matter. Excluded 
region (white at 90\% and Grey+white region at 66\% confidence level). Upper(lower) 
panel corresponds to normal(inverted) hierarchy.}
\label{fig:fig4}
\end{figure}
\begin{figure}[htb]
\includegraphics[width=6cm, height=4.5cm]{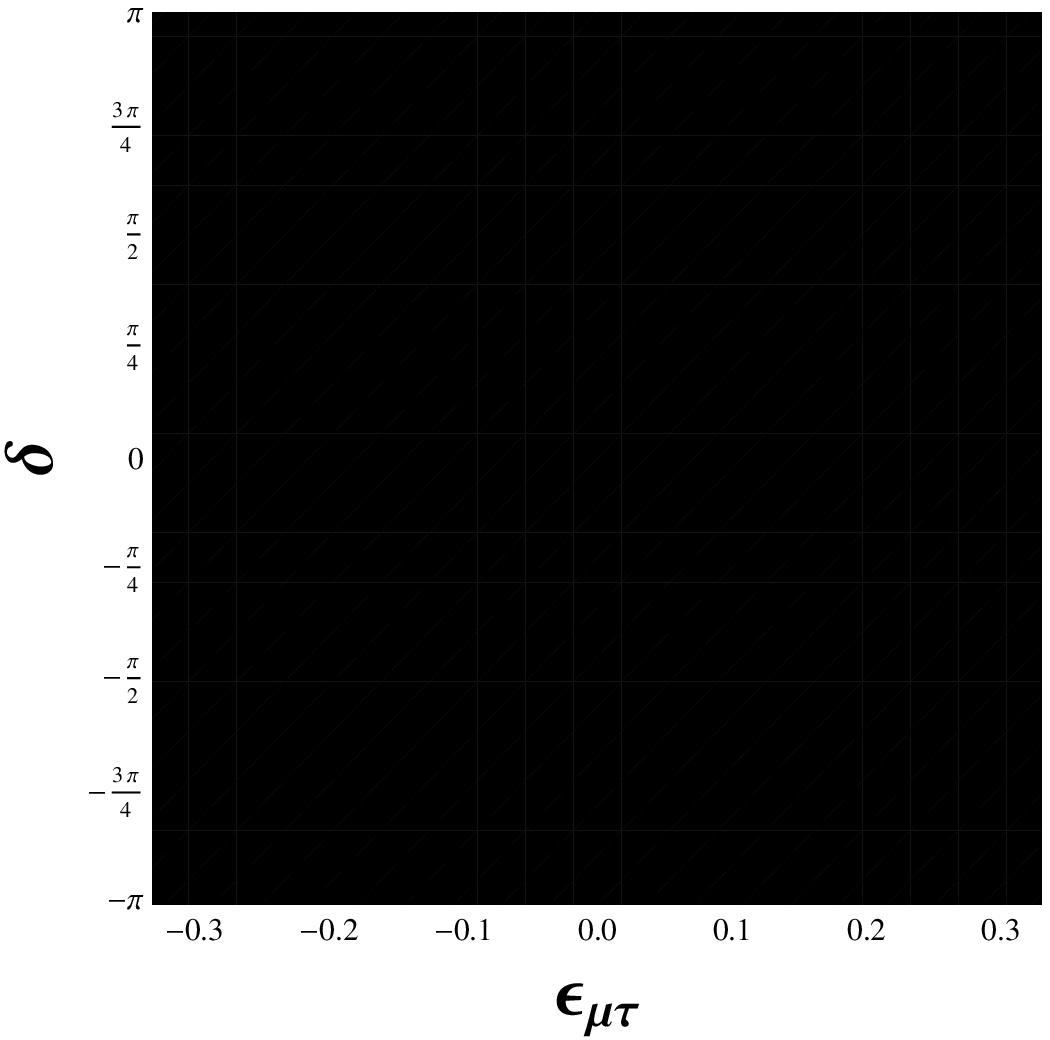}
\hspace{0.60cm}
\includegraphics[width=6cm, height=4.5cm]{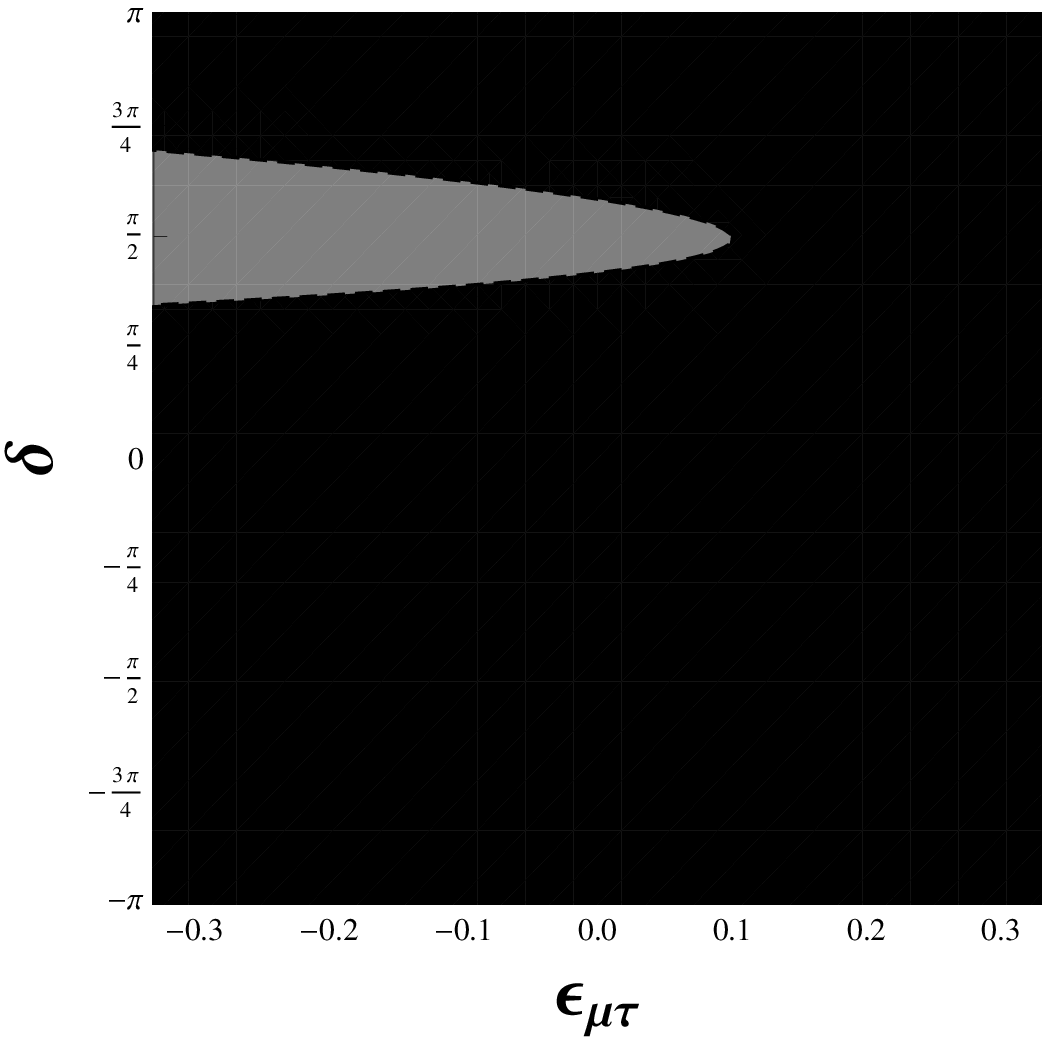}
\caption{Plot of $\epsilon_{\mu\tau} - \delta$ for  real NSI in matter. Excluded 
region (white at 90\% and Grey+white region at 66\% confidence level). Upper(lower) 
panel corresponds to normal(inverted) hierarchy.}
\label{fig:fig5}
\end{figure}

\begin{figure}[htb]
\includegraphics[width=6cm, height=4.5cm]{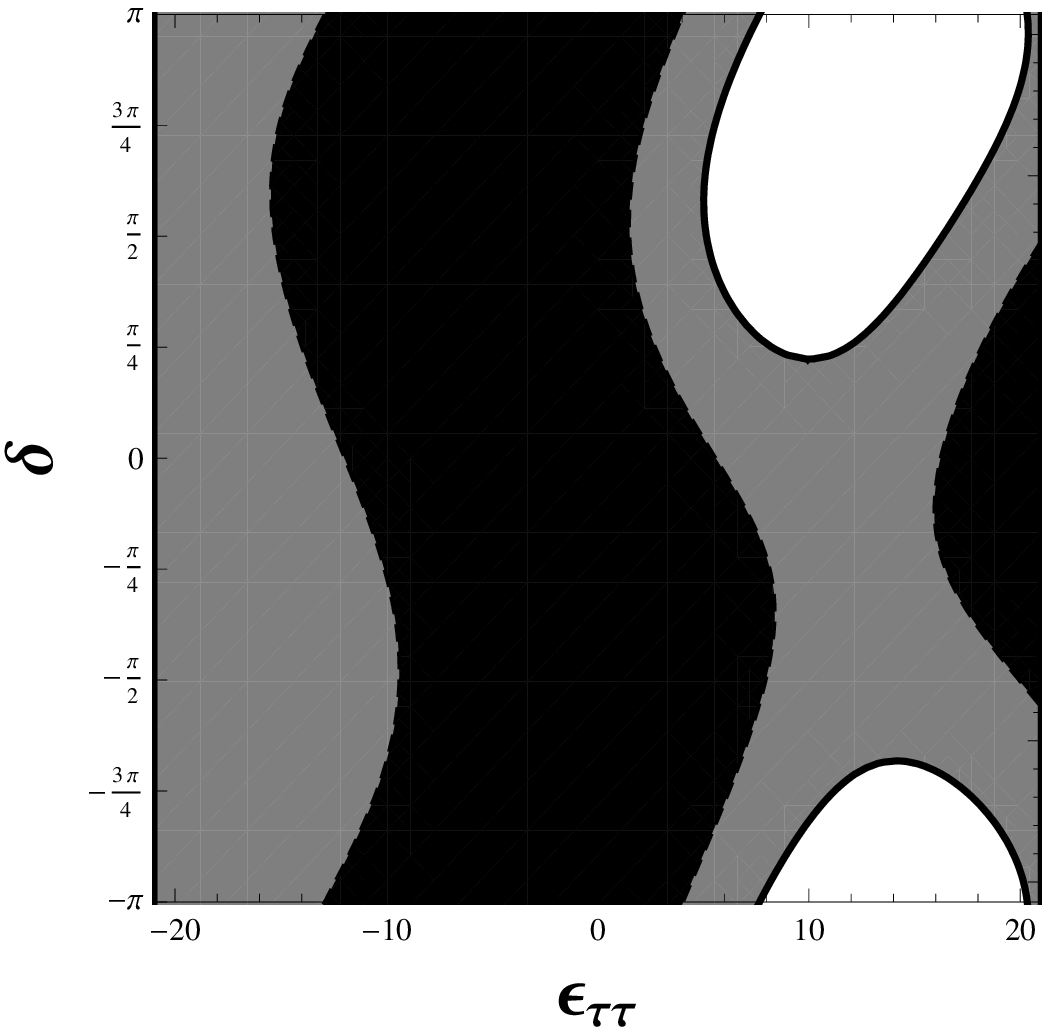}
\hspace{0.60cm}
\includegraphics[width=6cm, height=4.5cm]{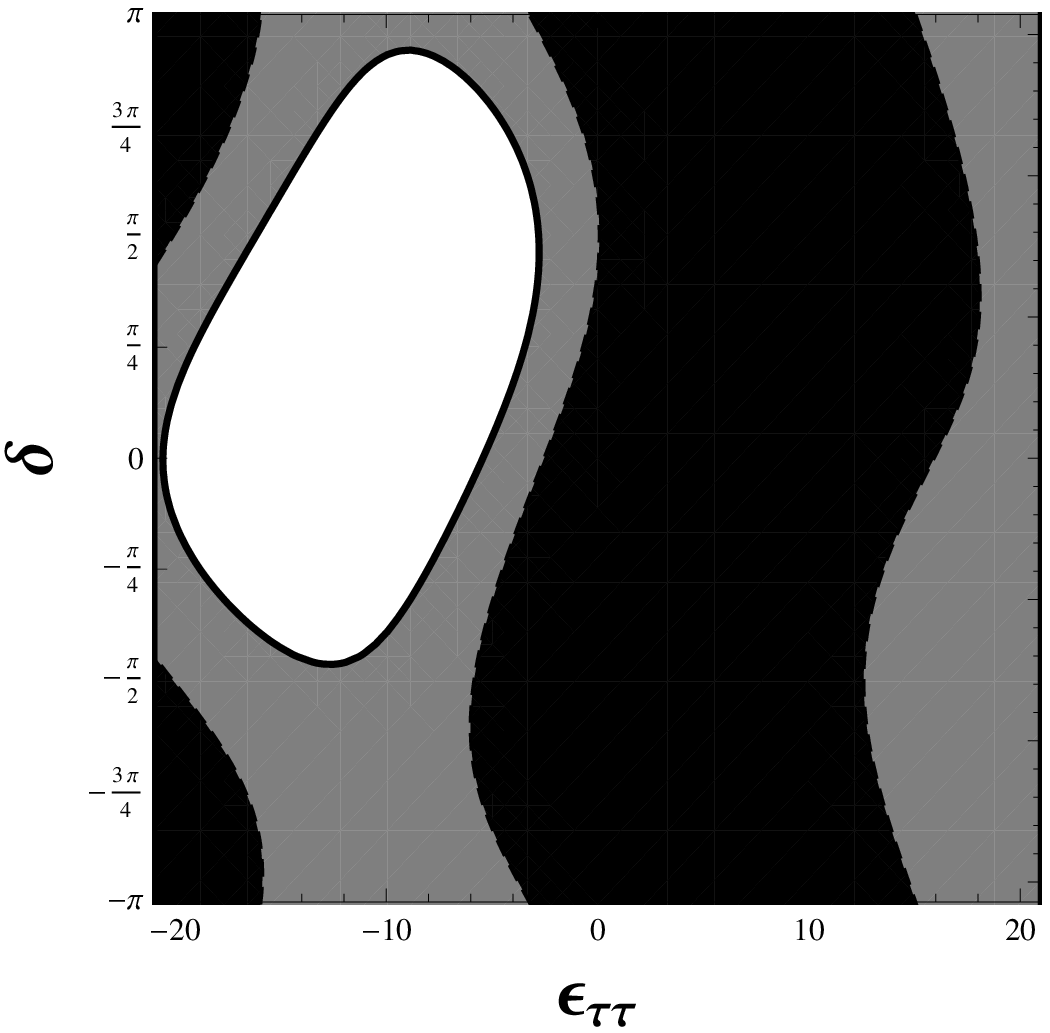}
\caption{Plot of $\epsilon_{\tau\tau} - \delta$ for  real NSI in matter. Excluded 
region (white at 90\% and Grey+white region at 66\% confidence level). Upper(lower) 
panel corresponds to normal(inverted) hierarchy.}
\label{fig:fig6}
\end{figure}

\noindent
paper giving constraints on NSIs.  In our numerical analysis we have considered even 
much higher values of NSI which has not been
considered in our perturbative approach but which are allowed after considering model 
independent constraints \cite{genbound-nsi}.
Some of the NSIs like  $\epsilon_{ee}$, $\epsilon_{\mu\mu}$, 
$\epsilon_{\tau\tau}$ (for propagation)  etc. do not appear in our expression of $P_{\nu_\mu \rightarrow \nu_e}$ in
 section IV and V as those have been assumed to be very small. In our numerical analysis however, we still have obtained some 
 constraints on those NSIs because of their presently allowed higher model-independent values as discussed later. 

For ultra-relativistic neutrinos, we have  
\begin{eqnarray}
 E_{k} \simeq E + \frac{m^2_{k}c^4}{2E}, \hspace{0.5cm} pc\simeq E, \hspace{0.5cm} ct\simeq x.
\end{eqnarray}
where $k =1,2,3$ correspond to mass eigenstates and $E_k$ are the eigenenergies. The neutrino energy $E$ is the average
energy after assuming the three momentum of different  components (1,2 and 3) to be equal.

Therefore, for the numerical analysis the relevant transition evolution equation for the 
flavor transition is
\begin{equation}
 i\hbar c \frac{d}{dx}S_{\beta \alpha}(x) = \sum_{\eta}\mathcal{H}_{\beta \eta}S_{\eta \alpha},
\end{equation}
with initial condition $S_{\beta \alpha}(0)=\delta_{\beta \alpha}$ and $\mathcal{H}$ is the 
total Hamiltonian comprising of standard matter interaction and NSI during propagation. 
However, here apart from NSI in propagation we want to include the source
and the detector NSI interaction. So, to implement that we apply the source and detector 
NSI matrices to the $S_{\alpha \beta}$ which 
is already mentioned in \eqref{eq:P-ansatz} as
\begin{eqnarray*}
\mathcal{A}_{\beta \alpha}  = \frac{1}{N^s_\alpha N^d_\beta}[ ( 1 + \epsilon^d )^T \,\, S \,\,
( 1 + \epsilon^s )^T ]_{\beta \alpha}.
\end{eqnarray*}
The  probability expression for the transition $\nu_{\alpha}\rightarrow \nu_{\beta}$ is
\begin{equation}
 P_{\nu^s_\alpha \rightarrow \nu^d_\beta}= |\mathcal{A}_{\beta \alpha}|^2.
\end{equation}
Although NSI considered at source, that at the detector and that during propagation are 
in general different, however, following \cite{Zhang} we have considered 
$\epsilon^s_{\alpha \beta}=\epsilon^{d*}_{\beta \alpha}$.

The recent T2K result \cite{T2K}  has obtained the constraint on $\delta-\sin^22\theta_{13}$ 
plane at  $90 \%$ confidence level based on the events in $\nu_\mu \rightarrow \nu_e$ 
transition in the baseline of 295 Km. Furthermore, Daya Bay reactor neutrino experiment 
has recently measured the neutrino mixing angle $\theta_{13}$ with 5.2 $\sigma$ confidence level for 
which $\sin^2 2\theta_{13}=0.092\pm 0.016 \pm 0.005 (syst)$.  Here, we analyze both these 
constraints considering real NSIs (one at a time) in propagation. Somewhat conservative bound 
on all NSIs at source and detector have been considered and taken to be of about $10^{-3}$. 
To tune with the experimental result of T2K and Daya Bay we shall use the same range of 
probability of oscillation as one obtains using the constraints on $\delta - \sin^2 2\theta_{13} $ 
given by T2K (without considering NSI) at different confidence level for normal and 
inverted hierarchies (allowing the variation of $\theta_{13}$ as in T2K paper). After that 
we shall fix $\theta_{13}$ at Daya Bay value and find out the allowed ranges in the 
parameters $\delta$ and different NSIs in matter (one at a time) subject to this constraint 
on the  probability of oscillation. For numerical analysis we use the following values 
as considered by T2K \cite{T2K} :
$\Delta m_{12}^2=7.6\times10^{-5} {\rm eV}^{2}$, 
$\Delta m_{23}^2=2.4\times 10^{-3} {\rm eV}^{2}$, $\sin^2 2\theta_{12}=0.8704$, $\sin^2 2\theta_{23}=1.0$, an average earth density 
$\rho=3.2 g/cm^{3}$  and central value of $\sin^22\theta_{13}=0.092$ as obtained 
from Daya Bay. One may note here that in presence of NSIs paticularly in matter 
(which are expected to be much larger than those NSIs at source and detector) the neutrino 
mixing parameters considered by T2K could change \cite{friedlandsolar,friedlandatmos}. However, 
it is found that only when several NSIs are considered simultaneously small changes occur 
in the best-fit values of these parameters. As for example, considering solar and KamLAND 
data the change in best-fit value of $\theta_{12}$ and $\Delta m_{12}^2$ can be seen
in figure 2 in \cite{friedlandsolar} and considering atmospheric and K2K data the 
change in best-fit value of $\theta_{23}$ and $\Delta m_{23}^2$ can be seen in figure 5  in
\cite{friedlandatmos}  after considering NSIs like $\epsilon_{ee}, \epsilon_{e\tau} $ 
and $\epsilon_{\tau\tau}$ simultaneously. However, in our analysis, we have obtained  
constraint on $\delta - $ NSI plane by considering one of the NSIs at a time for which 
the changes in these mixing parameters are expected to be small and in this simple analysis 
we use the values of mixing parameters as considered by T2K. For a very rigorous analysis 
in obtaining these mixing parameters in presence of NSIs, one
is required to fit the solar, KamLAND, atmospheric and K2K data simultaneously by 
considering NSIs one at a time or all together in the general three flavor neutrino mixing 
scenario which has not been done so far to the best of our knowledge.

In all the plots of NSI versus $\delta$ the $\epsilon_{\alpha\beta}$ corresponds to 
$\epsilon_{\alpha\beta}^m$ which are NSIs in matter during propagation. Dark shaded 
regions correspond to allowed region. T2K constraint
on $\delta - \sin^22 \theta_{13} $  plane at $66 \% $ confidence
level (C.L) together with Daya Bay result on $\theta_{13}$ correspond to excluded 
region (white+ grey) and only white excluded region correspond to the same T2K 
constraint at 90  \%  C.L. We have done the analysis on NSI constraints keeping in view 
the allowed range of NSIs for earth like matters as mentioned in \cite{genbound-nsi} 
and have considered those for real values. Upper (lower) panel in each plot correspond 
to normal (inverted) hierarchies. Out of various NSIs the significant constraints are obtained 
particularly for $\epsilon_{e\tau}^m$ and $\epsilon_{\tau\tau}^m$  for both the 
hierarchies of neutrino masses. Particularly for $\epsilon_{\mu\mu}$ and $\epsilon_{\mu\tau}$ 
no constraint can be obtained for normal hierarchy.

In Fig. \ref{fig:fig1} it is seen that in the upper panel for normal hierarchy the 
constraints on $\epsilon_{ee}$ can be obtained for negative values only for certain 
values of $\delta $ corresponding to T2K's  66\% confidence level result whereas for
inverted hierarchy the constraint is mainly on positive $\epsilon_{ee}$. However, 
for inverted hierarchy there is also excluded white region corresponding to T2K's 90\% 
confidence level result. In Fig. \ref{fig:fig2} the constraint on $\epsilon_{e\mu}$ is 
found mainly for positive values for both the hierarchies corresponding to T2K's 66\% 
confidence level result only. In Fig. \ref{fig:fig3} $\epsilon_{e\tau}$ is significantly 
constrained. For normal hierarchy the negative (positive) value could be 
constrained up to about -0.2(0.6) for certain values of $\delta$ corresponding to T2K's 66\% 
confidence level result. For inverted hierarchy such constraints are even more stringent 
and for certain values of $\delta$ all negative values could be excluded. In Fig. \ref{fig:fig4}
no constraint is obtained for $\epsilon_{\mu\mu}$ for normal hierachy. However, 
for inverted hierarchy around $\delta = \pi/2$ all values are excluded corresponding to 
T2K's 66\% confidence level result. In Fig. \ref{fig:fig5} for normal hierarchy no 
constraint is obtained on $\epsilon_{\mu\tau}$. For inverted hierarchy all negative 
values are excluded.  In Fig. \ref{fig:fig6} there are stringent constraints on
$\epsilon_{\tau\tau}$ particularly for positive values for normal hierarchy and negative 
values for inverted hierarchy corresponding to T2K's result at both 66\% and 90\% confidence 
level. Significant part of negative (positive) values of $\epsilon_{\tau\tau}$ are also 
excluded for normal(inverted) hierarchy corresponding to 66\% confidence level. 
Interestingly corresponding to T2K's 66\% confidence level result one may obtain some 
constraints on $\epsilon_{\tau\tau}$ independent of $\delta$ from Fig. \ref{fig:fig6}.

\section{Conclusion}
In this work, using perturbation theory, we obtain the probability of oscillation 
$P_{\nu_\mu \rightarrow \nu_e}$  (suitable for relatively short baseline of T2K and for large
 $\theta_{13}$ as evident from Daya Bay experiment) up to order $\alpha^2 ( \alpha \equiv 
 \frac{\Delta m_{21}^2}{\Delta m_{31}^2}$)  by considering NSIs at the source, 
detector as well as during propagation of neutrinos through matter. We have kept the 
standard matter interaction part in perturbed Hamiltonian which is appropriate for the 
baseline considered by T2K. In addition, we have considered two  cases, namely, 
$\epsilon_{\alpha\beta}^m\sim \epsilon\sim 0.03$ and $\epsilon_{\alpha\beta}^m\sim 
\sqrt{\epsilon}\sim 0.18$ -the latter corresponding to slightly larger NSI. In the expression 
of oscillation pobability one can see that a flavor transition takes place at the source 
even before the propagation of neutrinos due to NSI at source and detector - which  is the 
so-called zero distance effect. However, due to stringent constraints on these NSI 
parameters \cite{genbound-nsi} we have assumed all of them of about $10^{-3}$ in our 
numerical analysis. Although there are good model dependent bounds on NSI in matter 
(earth-like), these are not so strong if one likes to constrain them in a model independent 
way \cite{genbound-nsi}. In our numerical analysis, we have obtained constraints on various 
NSIs in matter in a model independent way from neutrino  oscillation experiments. 
Nevertheless, one may note as mentioned at the end of section V that it is difficult
to constrain very small NSIs in the relatively short baseline oscillation experiment like T2K. 
Recent Daya Bay result on mixing angle $\theta_{13}$ has helped us to give bound on NSI 
depending on only one so far unknown parameter $\delta$ in neutrino mixing matrix. Once 
one knows this phase from some short baseline neutrino oscillation experiments one may expect 
the better understanding about the possible strength of NSI. Depending on $\delta$ value 
significant constraint on $\epsilon_{e\tau}^m$ and $\epsilon_{\tau\tau}^m$ could be possible 
for both normal and inverted neutrino mass hierarchies. One may note here that in 
section IV $\epsilon_{\tau\tau}^m$ has not appeared in the expression of oscillation 
probability but still we have obtained significant constraint on it because of its very 
high presently allowed model independent values \cite{genbound-nsi}. Our studies indicate 
that while finding neutrino oscillation parameters it might be important to search for any 
possible evidence of NSIs even in the relatively shorter baseline neutrino oscillation 
experiments although longer ones are in general preferred. In the coming years the precision 
measurement of neutrino oscillation parameters and the NSI parameters in neutrino 
oscillation experiments could be challenging and could even show the evidence of NSIs.
%
\begin{acknowledgments} 
Two of us (R. A and A. D) acknowledge the hospitality of Indian Association for 
Cultivation of Science, Kolkata where this work was initiated. S. C and A. D like to 
thank Council of Scientific and Industrial Research, Government of India, for  junior 
research fellowship. S.C would also like to thank Kush Saha, Pradipta Ghosh and 
Subhadeep Mondal for helpful discussions and Tommy Ohlsson for his thoughtful insights 
which helped us greatly. S.R acknowledges the kind hospitality provided by the 
Helsinki Institute of Physics and the University of Helsinki, Finland and CERN theory 
division while this work was in progress. R.A and S.R would like to acknowledge the 
hospitality provided by the organizers of WHEPP-XII held at Mahabaleshwar, India where 
this work was completed.
\end{acknowledgments}



\end{document}